\documentclass{mekit19}
\usepackage{float,graphicx,amsmath,amsthm,amssymb,subcaption,bbold,lineno,hyperref}
\usepackage{multirow,bm,color}
\usepackage{cleveref}
\usepackage[square,numbers]{natbib}
\usepackage{autonum}
\modulolinenumbers[5]

\newcommand{\bom}{\boldsymbol{\omega}}

\newcommand{\trans}{^{\mathrm{T}}}

\newcommand{\R}{\mathbb{R}}
\newcommand{\Z}{\mathbb{Z}}

\newcommand{\T}{\mathbb{T}}

\newcommand{\bu}{\bm{u}}

\newcommand{\bx}{\bm{x}}
\newcommand{\X}{\bm{X}}
\newcommand{\be}{\bm{e}}

\newcommand{\bv}{\bm{v}}

\newcommand{\p}{\partial}
\renewcommand{\div}{{\rm{div}\,}}
\newcommand{\SB}{{\rm SB}}
\newcommand{\abs}[1]{\left\lvert #1 \right\rvert}

\newcommand{\mc}[1]{\mathcal{#1}}

\title{A SLENDER BODY MODEL FOR THIN RIGID FIBERS: VALIDATION AND COMPARISONS}

\author{Laurel Ohm$^{1}$, Benjamin K. Tapley$^{2}$, Helge I. Andersson$^{3}$,  Elena Celledoni$^{2}$, and Brynjulf Owren$^{2}$}

\heading{L. Ohm, B.K. Tapley, H.I. Andersson, E. Celledoni, and B. Owren}

\address{$^{1}$ School of Mathematics, University of Minnesota, 	 Minneapolis, MN 55455
	\and
	$^{2}$ Department of Mathematical Sciences, The Norwegian University of Science and Technology, 7491 Trondheim, Norway
	\and
	$^{3}$ Department of Energy and Process Engineering, The Norwegian University of Science and Technology, 7491 Trondheim, Norway 
	}


\abstract{In this paper we consider a computational model for the motion of thin, rigid fibers in viscous flows based on slender body theory. Slender body theory approximates the fluid velocity field about the fiber as the flow due to a distribution of singular solutions to the Stokes equations along the fiber centerline. The velocity of the fiber itself is often approximated by an asymptotic limit of this expression. Here we investigate the efficacy of simply evaluating the slender body velocity expression on a curve along the surface of the actual 3D fiber, rather than limiting to the fiber centerline. Doing so may yield an expression better suited for numerical simulation. We validate this model for two simple geometries, namely, thin ellipsoids and thin rings, and we compare the model to results in the literature for constant and shear flow. In the case of a fiber with straight centerline, the model coincides with the prolate spheroid model of Jeffery. For the thin torus, the computed force agrees with the asymptotically accurate values of Johnson and Wu and gives qualitatively similar dynamics to oblate spheroids of similar size and inertia.}

\begin{document} 

\section{Introduction}
Understanding the dynamics of particles immersed in viscous fluids is of importance in many areas of nature and industry. The first problem one encounters when simulating the dynamics of particles with complicated shapes is determining an appropriate model. As the forces and torques of arbitrarily shaped particles are not known in general, one must make a number of assumptions on the particle size and shape to accurately and cheaply specify the forces and torques on the particle. If the particle length scale is small (for example, smaller than the Kolmogorov scale in turbulent flows), the local fluid velocity can be accurately approximated by creeping Stokes flow and then the problem is amenable to a number of mathematical techniques that are available in the literature. One popular technique involves implementing slender body theories to model long and thin particles. An advantage of using slender body models is that they have the freedom to model flexible and arbitrarily shaped particles (with free ends or closed loops) provided that the particle is thin and the parametrization of centerline is known. The theoretical assumptions on which slender body models are based are also valid for long particles whose centerline lengths are comparable or extend beyond the limiting length scales of the fluid field. In particular, slender body theory has the potential to model particles that are longer than the Kolmogorov scale, where conventional models such as the Jeffery model for ellipsoids are not valid. This is a major advantage over current state-of-the-art particle simulations in, for example, \cite{mortensen2008dynamics,zhao2015rotation}. We also refer to \cite{voth2017anisotropic} and references therein for a review of other available models and methodologies for treating anisotropic particles in turbulent flows. 

In this article, we will consider a model based on slender body theory for rigid fibers that have either free ends or are closed loops. The purpose of this paper is primarily to provide a numerical validation of the proposed slender body model. For this reason, we will primarily focus on two simple geometries: long ellipsoids and thin rings (also referred to as thin tori). These geometries are chosen as there are verified ellipsoid and torus models available in the literature with well-studied dynamics, see for example \cite{zhang2001ellipsoidal,mortensen2008dynamics} for prolate ellipsoids and \cite{johnson1979hydromechanics} for thin torus models. This will serve as grounding for future work that will focus on more interesting and complex particle shapes (e.g., helical particles, complex closed loops or very long particles) in more complex flows (e.g., 3D numerical turbulence) that can be approached with more advanced numerical methods \cite{tapley2019novel,tapley2019computing}. Such studies could impact our understanding of the transport and deposition of microplastics in the ocean, since a large percentage of these microplastics are thin fibers \cite{martin2017deposition}.

The slender body approximation expresses the fluid velocity away from the fiber centerline as an integral of singular solutions to the Stokes equations along the fiber centerline. As such, the approximation itself is singular along the fiber centerline, and there exist various methods to obtain a limiting integral expression for the velocity of the slender body itself \cite{lighthill1976flagellar,keller1976slender,cortez2012slender}. For the purposes of particle simulations, we are primarily interested in solving for the forces and torques on the particle given a flow about the body. In the case of slender body theory, this involves inverting the limiting integral expression for the fiber velocity to find the force per unit length. Thus we need to be careful that the limiting expression is suitable for numerical inversion. In particular, we hope to avoid the high wavenumber instabilities that arise in some of the existing centerline expressions which require additional regularization to overcome. Often the methods for regularization lack a physical justification. 

Here we consider approximating the fiber velocity by simply evaluating the slender body fluid velocity expression on a curve along the actual slender body surface, away from the fiber centerline. Numerical evidence suggests that this method does not require further regularization to yield an invertible matrix equation for any discretization level or fiber centerline shape. We also show that our model agrees well with exact or asymptotically accurate expressions for the forces and torques on fibers with simple geometries in simple flows. 

The next section presents the mathematical theory for the slender body formalism, as well as a brief review of rigid body mechanics and spheroidal particle models. Section \ref{num} is dedicated to numerical experiments, and the final section is for conclusions. 

\section{Particle modeling}
We begin by reviewing the rigid body dynamics that are relevant to particle modeling. The theoretical basis for the slender body model is then presented for rigid free ended fibers and rigid closed loops. Finally, we present the Jeffery model for torques on an ellipsoid, which is used for comparison purposes. 

\subsection{Dynamics}
The angular momentum $\bm{m}$ of a rigid particle with torque $\bm{N}$ is governed by the ordinary differential equation
\begin{equation}\label{eq:rotation}
\dot{\bm{m}}=\bm{m}\times\bom+\bm{N},
\end{equation}
where $\bom = J^{-1} \bm{m}$ is the angular velocity and $J$ is the diagonal moment of inertia tensor. All the above quantities are given in the particle frame of reference. The particle orientation (with respect to a fixed inertial frame of reference) is specified using Euler parameters $q\in\mathbb{R}^4$ which satisfy the constraint $||q||_2=1$ and are determined by solving the ODE 
\begin{equation}\label{qode}
\dot{q} = \frac{1}{2} q\cdot w,
\end{equation}
where $w=(0,\bom\trans)\trans\in\mathbb{R}^4$ and $\cdot$ here denotes the Hamilton product of two quaternions \cite{goldstein2002classical}. A vector in the particle reference frame $\bx_p$ can be rotated to a vector in an inertial co-translating reference frame $\bx_T = Q \bx_p$ where $Q$ is the rotation matrix that is the image of $q$ under the Euler-Rodriguez map. We refer the reader to \cite{goldstein2002classical} for details on quaternion algebra and rigid body mechanics.
\subsection{Slender body theory}
We begin by describing the slender body geometries that will be considered in the free end and closed loop settings. To condense notation, we will use $\mc{I}$ to denote the interval $[-1/2,1/2]$ in the free end setting and the unit circle $\T=\R / \Z$ in the closed loop setting. We take $\X : \mc{I} \to \R^3$ to be the coordinates of an open or closed non-self-intersecting $C^2$ curve in $\R^3$, parameterized by arclength $s$. We let $\be_{\rm s}(s) = \frac{d\X}{ds}$ denote the unit tangent vector to $\X(s)$. The curve $\X(s)$ will be the centerline of the slender body, and we assume that all cross sections of the slender body are circular.  

\begin{figure}[!h]
	\centering
	\includegraphics[scale=0.66]{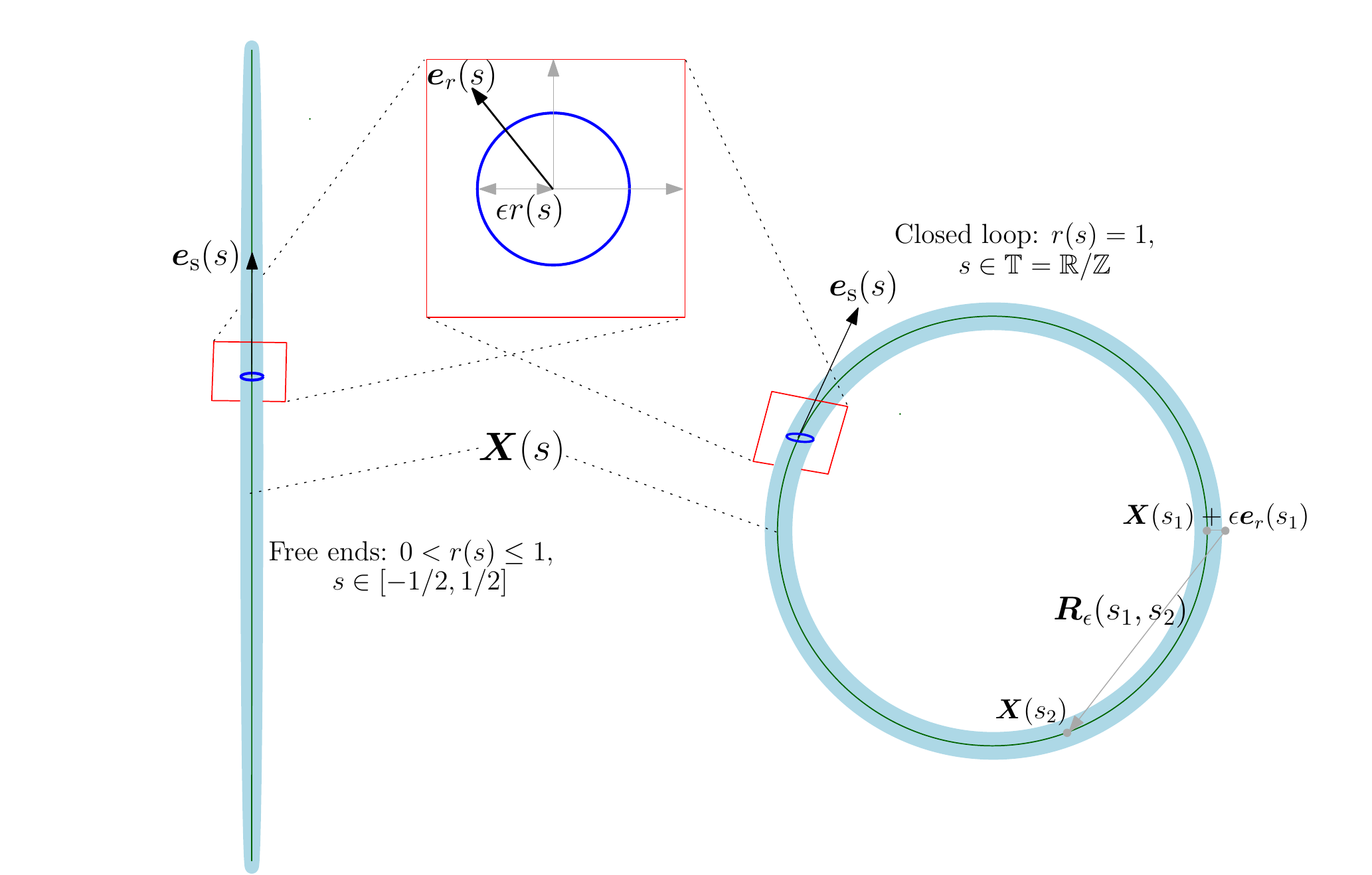}
	\caption{A depiction of the geometries under consideration in the free end and closed loop settings.}
	\label{fig:geometry}
\end{figure}

Let $0<\epsilon\ll 1$. In the closed loop setting, we consider fibers with uniform radius $\epsilon$ on each cross section. In the free end setting, we consider the actual endpoints of the fiber to be $\pm \sqrt{1/4+\epsilon^2}$ rather than $\pm 1/2$, and define a {\it radius function} $r\in C^2(-\sqrt{1/4+\epsilon^2},\sqrt{1/4+\epsilon^2})$ such that $0<r(s)\le1$ for each $s\in [-1/2,1/2]$, and $r(s)$ decays smoothly to zero at the fiber endpoints $\pm \sqrt{1/4+\epsilon^2}$. We will mostly be concerned with the prolate spheroid, for which we have 
\begin{equation}\label{prolate}
r(s) = \frac{1}{(\frac{1}{4}+\epsilon^2)^{1/2}}\bigg(\frac{1}{4}+\epsilon^2-s^2 \bigg)^{1/2}.
\end{equation}
Notice that the interval $[-1/2,1/2]$ extends from focus to focus of this prolate spheroid, and that $r=\mc{O}(\epsilon)$ at $s=\pm \frac{1}{2}$ (see figure \ref{fig:geometry}). In numerical applications, we will also briefly consider the case of a free end fiber with uniform radius (except for hemispherical caps at the fiber endpoints -- see section \ref{free_valid}), but we note that the slender body approximation is better suited for the prolate spheroid. Throughout this paper, for the sake of conciseness, we will often write one expression to encompass both the free end and closed loop settings, in which case we note that in the closed loop setting we define $r(s)= 1$ for each $s\in \T$. 

The idea behind slender body theory is to approximate the fluid velocity about the fiber as the Stokes flow due to a one-dimensional curve of point forces in $\R^3$. The basic theory originated with Hancock \cite{hancock1953self}, Cox \cite{cox1970motion}, and Batchelor \cite{batchelor1970slender} with later improvements by Keller and Rubinow \cite{keller1976slender} and Johnson \cite{johnson1980improved}. Here we will consider specifically the slender body theory of Johnson, which was further studied by G\"otz \cite{gotz2000interactions} and Tornberg and Shelley \cite{tornberg2004simulating}. Let $\bu_0(\bx,t)$ denote the (known) velocity of the fluid in the absence of the fiber at time $t$, and let $\mu$ denote the viscosity of the fluid. The classical slender body approximation $\bu^\SB(\bx,t)$ to the fluid velocity at any point $\bx$ away from the fiber centerline $\X(s,t)$ is then given by 
\begin{align}
8\pi\mu \big(\bu^{\SB}(\bx,t) - \bu_0(\bx,t) \big) &=-\int_{\mc{I}} \bigg( \mc{S}(\bm{R})+\frac{\epsilon^2r^2(s')}{2}\mc{D}(\bm{R}) \bigg)\bm{f}(s',t) \, ds', \quad 
\bm{R}=\bm{x}-\bm{X}(s',t); \\
\label{stokes_SB}
\mc{S}(\bm{R})&=\frac{{\bf I}}{\abs{\bm{R}}}+\frac{\bm{R}\bm{R}^{\rm T}}{\abs{\bm{R}}^3}, \; 
\mc{D}(\bm{R})=\frac{{\bf I}}{\abs{\bm{R}}^3}-\frac{3\bm{R}\bm{R}^{\rm T}}{\abs{\bm{R}}^5}.
\end{align}
Here $\frac{1}{8\pi\mu}\mc{S}(\bm{R})$ is the Stokeslet, the free space Green's function for the Stokes equations in $\R^3$, and $\frac{1}{8\pi\mu}\mc{D}(\bm{R})=\frac{1}{16\pi\mu}\Delta\mc{S}(\bm{R})$ is the doublet, a higher order correction to the velocity approximation. The force density $\bm{f}(s,t)$ is here considered as the force per unit length exerted by the fluid on the body. The sign convention is opposite if we instead consider $\bm{f}$ to be the force exerted by the body on the fluid. Note that in the free end case, this force density is only distributed between the generalized foci of the slender body ($s=\pm 1/2$) rather than between the actual endpoints of the fiber.  \\

In the stationary setting, Mori et al. in \cite{closed_loop} (closed loop case) and \cite{free_ends} (free end case) prove a rigorous error bound for the difference between the velocity field given by \eqref{stokes_SB} and the velocity field around a three-dimensional flexible rod satisfying a well-posed {\it slender body PDE}. In particular, for the closed loop, given a force density $\bm{f}\in C^1(\T)$, the difference between $\bu^{\SB}$ and the PDE solution exterior to the slender body is bounded by an expression proportional to $\epsilon\abs{\log\epsilon}$. In the free end case, given a force density $\bm{f}\in C^1(-1/2,1/2)$ which decays like a spheroid at the fiber endpoints ($\bm{f}(s)\sim \sqrt{1/4 - s^2}$ as $s\to \pm 1/2$), the difference between the free end slender body approximation $\bu^{\SB}$ and the well-posed PDE solution of \cite{free_ends} is similarly bounded by an expression proportional to $\epsilon\abs{\log\epsilon}$. Thus the Stokeslet/doublet expression \eqref{stokes_SB} is quantitatively a good approximation of the flow field around a slender body. \\

To approximate the velocity of the slender body itself, we would like to use \eqref{stokes_SB} to obtain an expression for the relative velocity of the fiber centerline $\frac{\p\X(s,t)}{\p t}$ depending only on the arclength parameter $s$ and time $t$. In the case of a rigid fiber, given the velocity $\frac{\p\X(s,t)}{\p t}=\bv+\bm{\omega}\times \X(s,t)$, $\bv,\bm{\omega}\in\R^3$, of the filament centerline, we would like to then be able to invert the centerline velocity expression to solve for the force density $\bm{f}(s,t)$ along the fiber. We use this $\bm{f}(s,t)$ to compute the total force $\bm{F}(t)$ and torque $\bm{N}(t)$ exerted on the body as 
\begin{equation}\label{slender body theory_FandT}
 \int_{\mc{I}} \bm{f}(s,t) \, ds = \bm{F}(t), \quad \int_{\mc{I}} \X(s,t)\times \bm{f}(s,t) \, ds = \bm{N}(t).
 \end{equation}
Since the expression \eqref{stokes_SB} is singular at $\bx=\X(s,t)$, deriving a limiting expression for the fiber centerline must be done carefully. There are various ways to use \eqref{stokes_SB} to obtain a centerline expression depending on $s$ only, including the methods of Lighthill \cite{lighthill1976flagellar}, Keller and Rubinow \cite{keller1976slender}, and the method of regularized Stokeslets \cite{bouzarth2011modeling,cortez2005method,cortez2012slender}. Each method expresses the velocity of the slender body centerline $\frac{\p\X(s,t)}{\p t}$ as an integral operator acting on the force density $\bm{f}(s,t)$. A brief overview of these methods is given in appendix \ref{append}. 

Because solving for the force density $\bm{f}(s,t)$ given $\frac{\p\X(s,t)}{\p t}$ involves inverting an integral operator at each time step, we need to take particular care that the operator -- at least when discretized -- is suitable for inversion. In particular, we need to avoid the high wavenumber instabilities that limit discretization of the integral operator and hinder some of the asymptotic methods described in appendix \ref{append}. At the same time, we would like the centerline expression to have a clear physical meaning and connection to the Stokeslet/doublet expression \eqref{stokes_SB}.  

Thus we will use the following expression to approximate the velocity $\frac{\p\X(s,t)}{\p t}$ of the slender body itself. Taking $\be_r(s,t)$ to be a particular unit vector normal to $\X(s,t)$ (we will discuss the choice of $\be_r$ later), we essentially evaluate \eqref{stokes_SB} at $\bx=\X(s,t) + \epsilon r(s)\be_r(s,t)$, a curve along the actual surface of the slender body. For $\mc{S}$, $\mc{D}$ as in \eqref{stokes_SB}, we have
\begin{align}\label{SB_new}
8\pi\mu \bigg(\frac{\p\X}{\p t} &- \bu_0(\X(s,t),t) \bigg) = - \int_{\mc{I}}\bigg(\mc{S}_\epsilon (s,s',t)+ \frac{\epsilon^2r^2(s')}{2} \mc{D}_\epsilon(s,s',t)  \bigg) \bm{f}(s',t) \, ds' ;  \\
\label{SD0}
\mc{S}_\epsilon &= \mc{S}(\bm{R}_\epsilon(s,s',t)) - \frac{\epsilon^2 r^2 \be_r\be_r^{\rm T}}{\abs{\bm{R}_\epsilon(s,s',t)}^3}, \quad \mc{D}_\epsilon = \mc{D}(\bm{R}_\epsilon(s,s',t)) + \frac{3\epsilon^2 r^2 \be_r\be_r^{\rm T}}{\abs{\bm{R}_\epsilon(s,s',t)}^5}, \\
&\bm{R}_\epsilon(s,s',t)=\X(s,t) - \X(s',t) + \epsilon r(s)\be_r(s,t).
\end{align}
Here we are relying on the fact that for any point $\bx$ on the actual fiber surface, the expression \eqref{stokes_SB} for $\bu^{\SB}(\bx)$ is designed to depend only on arclength $s$ to leading order in $\epsilon$ -- in particular, on each cross section of the slender body, the angular dependence about the fiber centerline is only $\mc{O}(\epsilon\log\epsilon)$ (see \cite{closed_loop}, proposition 3.9, and \cite{free_ends}, proposition 3.11). This is because the leading order angular-dependent terms (the $\epsilon^2 r^2 \be_r\be_r^{\rm T}$ term in both the Stokeslet and the doublet, which is $\mc{O}(1)$ at $s=s'$) cancel each other asymptotically to order $\epsilon\log(\epsilon)$ (see estimates 3.62 and 3.65 in \cite{closed_loop} and estimates 3.40 and 3.43 in \cite{free_ends}). We therefore eliminate these two terms from the formulation \eqref{SB_new}, in part due to this cancellation and in part because their omission appears to improve the stability of the discretized integral operator \eqref{SB_new} when $n$, the number of discretization points, is large. This apparent improvement in stability merits further study in future work. 

Thus to approximate the velocity of the fiber centerline, we evaluate \eqref{stokes_SB} on the actual slender body surface along a normal vector $\be_r(s,t)\in C^2(\mc{I})$ extending from $\X(s,t)$ but cancel the $\epsilon^2 r^2 \be_r\be_r^{\rm T}$ terms that would otherwise appear. Note that the choice of normal vector $\be_r$ is somewhat arbitrary, and does have an $\mc{O}(\epsilon\log\epsilon)$ effect on the resulting approximation. These effects can and should be studied further in future work. However, we use this normal vector as a physically meaningful means of avoiding the high wavenumber instabilities that appear in other asymptotic methods (see appendix \ref{append}). Numerical evidence suggests that the discretized centerline equation \eqref{SB_new} yields a matrix equation that is solvable for $\bm{f}(s,t)$ given $\frac{\p\X(s,t)}{\p t}$, as all eigenvalues of the matrix are positive even for very large $n$.  This is not necessarily the case for some of the other centerline equations (again, see appendix \ref{append}) unless additional regularizations are added, which may affect the physical meaning of the equations. The possibility of resolving very fine scales along the length of the fiber is desirable especially when dealing with turbulent flows.  

\subsection{Spheroid model}
The above slender body model is valid for arbitrary parameterizations of the centerline $\bm{X}(s,t)$ and a wide choice of radius functions. However, to validate the model we will focus on a simple case where the centerline is a straight line and the radius function corresponds to an ellipsoid. In this case the torques have a known expression due to Jeffery \cite{jeffery1922motion} and the motion of such a particle in simple flows is well-known \cite{challabotla2015rotational,mao2014motion} which makes this choice of geometry a perfect arena for model validation. We will now briefly review some theory related to spheroids immersed in viscous fluids.

An axisymmetric spheroid in the particle frame is given by 
\begin{equation}\label{eqn:spheroid}
\frac{x^2}{a^2}+\frac{y^2}{a^2}+\frac{z^2}{b^2} = 1,
\end{equation}
where $a$ and $b$ are the distinct semi-axis lengths. The particle shape is characterized by the dimensionless aspect ratio $\lambda = b/a>0$, which distinguishes between spherical ($\lambda = 1$), prolate ($\lambda > 1$) and oblate ($\lambda < 1$) particles (the latter two shapes are also called as rods and disks). In the case of a slender prolate spheroid, we take $a=\epsilon$. The axisymmetric moment of inertia tensor for a spheroid in the body frame is
\begin{equation}
J = ma^2\mathrm{diag}\left(\frac{(1+\lambda^2)}{5},\frac{(1+\lambda^2)}{5},\frac{2}{5}\right),
\end{equation}
where $m=\frac{4}{3}\pi \lambda a^3 \rho_p$ is the particle mass and $\rho_p$ is the particle density. Jeffery \cite{jeffery1922motion} calculated the torque $\mathbf{N}$ of an ellipsoid in creeping Stokes flow, which in the above axisymmetric case reads
\begin{align}
N_x = & \frac{16\pi \lambda\mu a^3}{3(\beta_0+\lambda^2\gamma_0)}\left[(1-\lambda^2)S_{yz}+(1+\lambda^2)(\Omega_x-\omega_x)\right],\label{eq:JT1} \\ 	
N_y = & \frac{16\pi \lambda\mu a^3}{3(\alpha_0+\lambda^2\gamma_0)}\left[(\lambda^2-1)S_{zx}+(1+\lambda^2)(\Omega_y-\omega_y)\right], \label{Jeffery} \\ 
N_z = & \frac{32\pi \lambda\mu a^3}{3(\alpha_0+\beta_0)}(\Omega_z-\omega_z),\label{eq:JT3}
\end{align}
where $S_{ij}=\frac{1}{2}\left(\frac{\partial u_i}{\partial x_j}+\frac{\partial u_j}{\partial x_i}\right)$ is the fluid shear tensor and $\boldsymbol{\Omega} = \frac{1}{2}\nabla\times \bu$ is the fluid rotation, both taking constant values in shear flow. The values $\alpha_0$, $\beta_0$ and $\gamma_0$ are $\lambda$-dependent parameters that were calculated in \cite{gallily1979orderly}. 

There are a number of distinctions to make between this model and the slender body model. First, Jeffery assumes that the particle is small enough that the fluid Jacobian $\nabla \bu$ is constant across the volume of the spheroid. In shear flow, $\nabla \bu$ is constant everywhere, hence this assumption is true and the model validity is independent of the size of the particle. However, in more complex flows such as turbulence, the Jeffery model is only valid for $a,b<<\eta$ for Kolmogorov length $\eta$. On the other hand, the slender model requires only that the maximal cross sectional radius $\epsilon<<\eta$ to be valid. Hence, the slender body model is valid for particles with lengths larger than $\eta$ whilst satisfying the Stokes flow assumptions.  Second, the Jeffery torque depends on the fluid velocity derivatives only, while the slender body model derives the torques from the velocity field along the centerline. Because of this, we cannot expect the models to coincide when the particle is aligned exactly in the shear plane (i.e., the plane where $\bu=\bm{0}$ but $\frac{\partial u_j}{\partial x_i}\ne 0$). 



\section{Numerical experiments}\label{num}
This section presents numerical results for the slender body model and comparisons with other similar models. We begin with a validation of the slender body expression \eqref{SB_new} by comparing the total force $\bm{F}$ given by inverting \eqref{SB_new} for a stationary slender body velocity with the exact expression for the Stokes drag on a particular object (when available) or with an expression valid asymptotically as $\epsilon\to 0$. We consider the slender prolate spheroid (section \ref{free_valid}; exact expression given by Chwang and Wu \cite{chwang1975hydromechanics}), the straight, uniform cylinder with hemispherical endpoints (section \ref{free_valid}; asymptotic expression given by Keller and Rubinow \cite{keller1976slender}), and the slender torus (section \ref{loop_valid}; asymptotic expression given by Johnson and Wu \cite{johnson1979hydromechanics}). In each case we expect $\mc{O}(\epsilon\log\epsilon)$ agreement between the force $\bm{F}$ computed using \eqref{SB_new} and the exact or asymptotically accurate expressions; however, we find that this trend is clearly visible only in the closed loop setting. We then examine the rotational dynamics of a prolate spheroid in shear flow using expression \eqref{SB_new} and compare it with the Jeffery model for ellipsoids \cite{jeffery1922motion}. We look at the dynamics of the two models for a range of aspect ratios and orientations and then explore the effect of the discretization parameter on the periodic Jeffery orbits. We finally compare the dynamics of thin rings to oblate spheroids for a range of fluid viscosities. 

\subsection{Computational considerations}
In many applications, one needs to simulate the dynamics of thousands or millions of particles; hence computational cost plays a role in determining the model choice. One thing to consider is that the slender body model involves inverting a $3n\times 3n$ matrix at each time step, where $n$ is the user-defined discretization parameter that arises from discretizing the integral in equation \eqref{SB_new}. On the other hand, the Jeffery model requires an accurate approximation of the fluid Jacobian at the location of the particle center of mass, while the slender body model only requires the fluid velocity values at the $n$ locations on its centerline. When the fluid velocity is defined at discrete locations in space, such as in direct numerical simulations of turbulent flows, the Jeffery model is faced with the problem of approximating the fluid Jacobian at the location of the particle center of mass, which is more costly than just interpolating the velocity field. In practice, however, one should use the Jeffery model when computing dynamics of small, thin ellipsoids when possible and the slender body model for more complicated shapes or longer particles. As the purpose of this article is focused on the theoretical and numerical validation of the slender body model, computational cost and numerical methods will be left for future work. 

\subsection{Free ended fibers in constant flow}\label{free_valid}
We validate the free end formulation of \eqref{SB_new} in the case of a slender body with straight centerline $\X(s)=s\be_x$, $s\in [-1/2,1/2]$, aligned with the $x$-axis. Here we will consider both the slender prolate spheroid with radius function $r(s)$ as in \eqref{prolate} and a slender cylinder with hemispherical caps at the fiber endpoints. In both cases, we take the actual filament length to be $2\sqrt{1/4+\epsilon^2}$, but distribute the force density $\bm{f}(s)$ only along $[-1/2,1/2]$. As in the closed loop setting, we use \eqref{SB_new} to calculate the drag force $\bm{F}$ on the slender body as it translates with unit speed, and compare this $\bm{F}$ to either exact or asymptotically accurate expressions for the Stokes drag on a prolate spheroid or cylinder. In both cases we will use the unit normal vector $\be_r(s) = \cos(2\pi s)\be_y +\sin(2\pi s)\be_z$, which rotates once in the $yz$-plane perpendicular to $\X(s)= s\be_x$ for $s\in [-1/2,1/2]$. This normal vector is chosen because it represents a sort of average normal direction along the length of the filament.

In the free end setting, we also need to make sure that the computed force density $\bm{f}(s)$ is decaying sufficiently rapidly at the fiber endpoints to ensure that the solution makes sense physically. The inclusion of the decaying radius function $r(s)$ in the slender body velocity expression \eqref{SB_new} ensures this decay by making the integral kernel very large near the fiber endpoints.

In the case of a prolate spheroid, we can actually compare the total force $\bm{F}$ given by \eqref{SB_new} to the analytical expression for Stokes drag on a spheroid calculated by Chwang and Wu \cite{chwang1975hydromechanics} (see table \ref{validate2}). We consider the drag force on a slender prolate spheroid translating with unit speed in either the $y$-direction (perpendicular to the semi-major axis) or the $x$-direction (parallel to the semi-major axis). In all cases, the integral term of \eqref{SB_new} is discretized using the trapezoidal rule with uniform discretization along the filament centerline. We use $n=2/\epsilon$ discretization points. 

\begin{table}[!h]
	\centering
	\begin{tabular}{ | c | c | c || c | c | c |} 
		\hline
		& \multicolumn{2}{|c||}{ $\bm{F}\cdot\be_y$ for $\bu=\be_y$} & \multicolumn{2}{|c|}{$\bm{F}\cdot\be_x$ for $\bu=\be_x$} & \\
		\hline
		$\epsilon$  & Expression \eqref{SB_new} & Chwang-Wu & Expression \eqref{SB_new} & Chwang-Wu & $\epsilon\abs{\log\epsilon}$ \\ 
		\hline
		0.01 & -2.4498  & -2.4618 & -1.5245 & -1.5302  & 0.0461\\   
		0.005 & -2.1579  & -2.1673  & -1.3051 & -1.3094  & 0.0265\\ 
		0.0025 & -1.9281 & -1.9358  & -1.1408 & -1.1442  & 0.0150\\ 
		0.00125 & -1.7426 & -1.7491 & -1.0133 & -1.0159 & 0.0084 \\
		\hline
	\end{tabular}
	\caption{Comparison of the computed (via expression \eqref{SB_new}) and exact (from Chwang and Wu \cite{chwang1975hydromechanics}) Stokes drag force $\bm{F}$ on a slender prolate spheroid of length $2\sqrt{1/4+\epsilon^2}$ with semi-major axis aligned with the $x$-axis. Columns 2 and 3 compare the $y$-component of $\bm{F}$ for a spheroid translating with unit speed in the $y$-direction, while columns 4 and 5 compare the $x$-component of $\bm{F}$ for translation in the $x$-direction. Note that for both directions, the force difference decreases with $\epsilon$, but not quite at the expected $\epsilon\log\epsilon$ rate. } \label{validate2}
\end{table}

\begin{figure}[!h]
	\centering
	\includegraphics[scale=0.35]{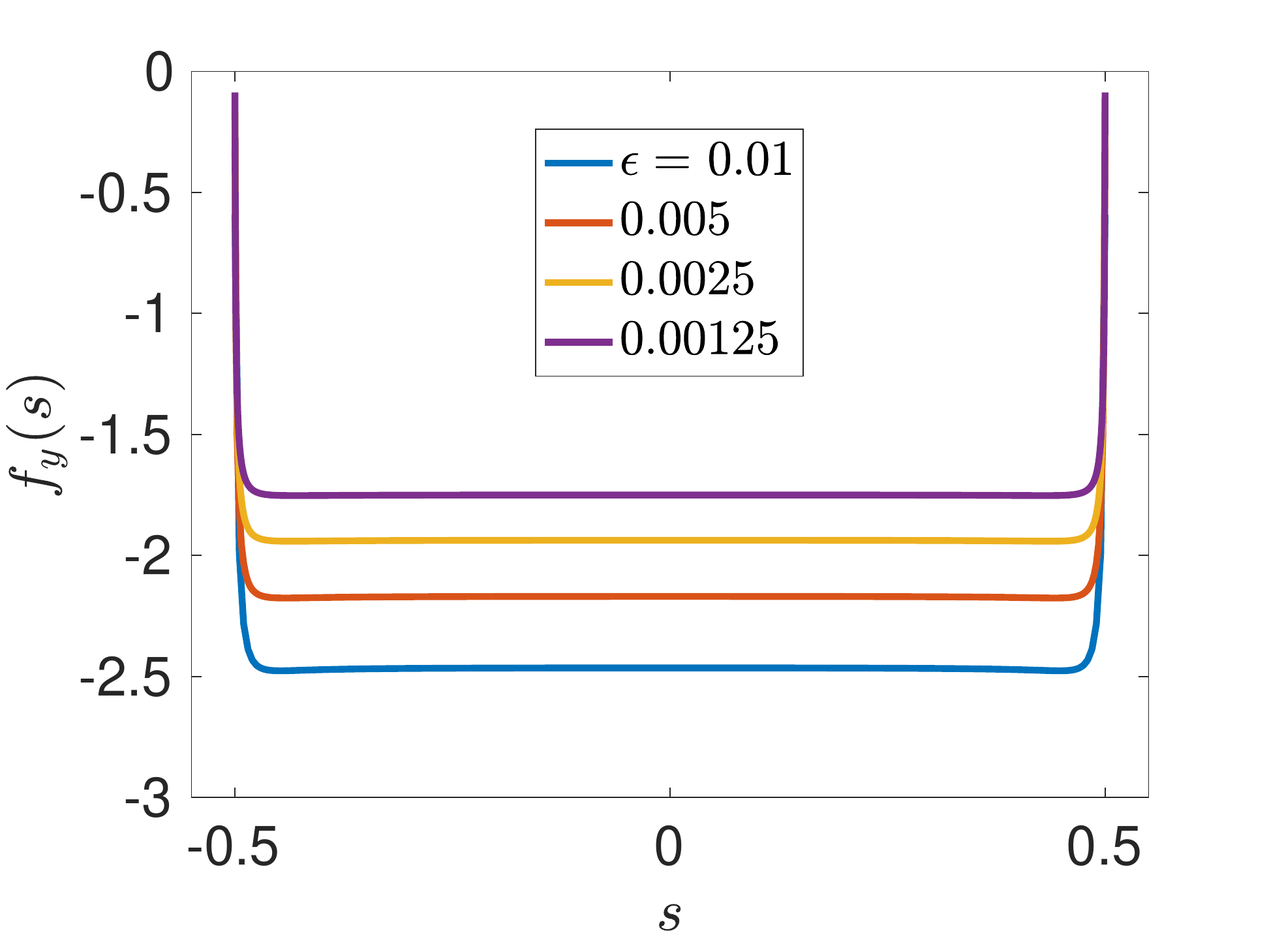}
	\includegraphics[scale=0.35]{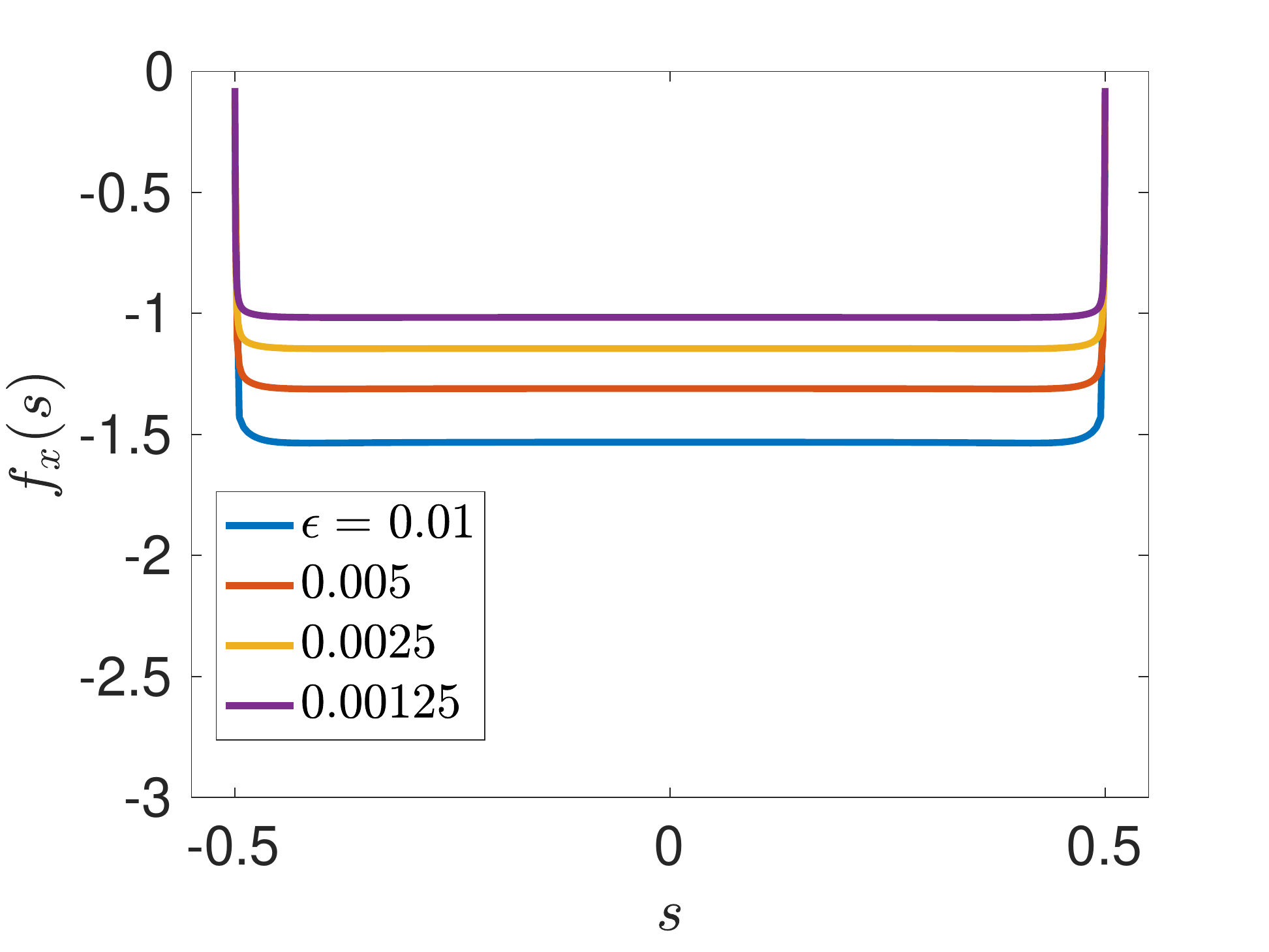}
	\caption{Force per unit length $\bm{f}(s)$, $s\in [-1/2,1/2]$, along the prolate spheroid with semi-major axis aligned with the $x$-axis.  The left figure shows the $y$-component of the force density for the cylinder translating with unit speed in the $y$-direction, while the right figure shows the $x$-component of the force density for the cylinder translating in the $x$-direction. Note that in both flows the force density $\bm{f}(s)$ decays to near zero at $s=\pm1/2$, as expected. }
	\label{fig:prolate}
\end{figure}

We also look at a plot of the computed force per unit length $\bm{f}(s)$ along the filament (figure \ref{fig:prolate}) to verify that the force density makes sense physically. 

From figure \ref{fig:prolate}, we can see that the force density $\bm{f}(s)$ decays rapidly as $s\to \pm 1/2$, but does not vanish identically at $\abs{s}=1/2$. However, it should be noted that in \cite{free_ends}, we are given the force density $\bm{f}(s)$, $s\in [-1/2,1/2]$, and use it to solve for the corresponding slender body velocity. In that case, the force must vanish identically at $\pm 1/2$ to yield a unique velocity. Since in this case we are using the fiber velocity to solve for the force density, it appears that what we are doing instead here is ignoring a certain (small) amount of force contribution from the very ends of the fiber (between $1/2\le \abs{s} \le \sqrt{1/4+\epsilon^2}$). Whether or not this is a good approximation is unclear -- it is possible that the same force density could result from flows that differ slightly at the actual fiber endpoints. However, it appears that because $\bm{f}(s)$ decays so rapidly at $s=\pm1/2$, any force contribution beyond this would be negligible. This may indicate that sufficient decay in the slender body radius toward the endpoints of the fiber ensures that the endpoints (beyond $\abs{s}=1/2$) are not contributing a significant amount to the total force and thus can be safely ignored.  


To test the formulation \eqref{SB_new} for a different choice of radius function $r(s)$, we next consider the drag force on a straight cylinder with uniform radius everywhere along its length except for hemispherical caps at the fiber endpoints. In particular, we take the cylinder to be the same length as the prolate spheroid (actual fiber endpoints at $s= \pm \sqrt{1/4+\epsilon^2}$) with a radius that decays smoothly to zero at the endpoint via a hemispherical cap of radius $\epsilon$ centered at $d_\epsilon=\sqrt{1/4 + \epsilon^2} - \epsilon$:
\begin{equation}\label{cylinder_rad}
\begin{aligned}
 \epsilon r(s) &= \begin{cases}
\epsilon, & -d_\epsilon \le s \le d_\epsilon \\
\sqrt{\epsilon^2-(s + d_\epsilon)^2}, & s < -d_\epsilon\\
\sqrt{\epsilon^2-(s - d_\epsilon)^2}, & s> d_\epsilon
\end{cases} \\
d_\epsilon &:= \sqrt{1/4 + \epsilon^2} - \epsilon.
\end{aligned}
\end{equation}

As in the case of the prolate spheroid, we distribute the force density $\bm{f}(s)$ along the interval $[-1/2,1/2]$. Using \eqref{SB_new} to find $\bm{F}$ in the same way as in the case of the prolate spheroid, we compare the resulting drag force with the asymptotic expression derived by Keller and Rubinow \cite{keller1976slender} in table \ref{validate3}.

\begin{table}[!h]
	\centering
	\begin{tabular}{ | c | c | c || c | c | c |} 
		\hline
		& \multicolumn{2}{|c||}{ $\bm{F}\cdot\be_y$ for $\bu=\be_y$} & \multicolumn{2}{|c|}{$\bm{F}\cdot\be_x$ for $\bu=\be_x$} & \\
		\hline
		$\epsilon$  & Eqn \eqref{SB_new} & Keller-Rubinow & Eqn \eqref{SB_new} & Keller-Rubinow  & $\epsilon\abs{\log\epsilon}$ \\ 
		\hline
		0.01 & -2.6433 & -2.6401 & -1.6864 & -1.6712 & 0.0461 \\ 
		0.005 & -2.3085 & -2.3024 & -1.4216 & -1.4094 & 0.0265 \\ 
		0.0025 & -2.0472 & -2.0417 & -1.2274 & -1.2189 & 0.0150 \\ 
		0.00125 & -1.8384 & -1.8342 & -1.0796 & -1.0738 & 0.0084 \\
		\hline
	\end{tabular}
	\caption{Comparison of the computed (via expression \eqref{SB_new}) and asymptotic (from Keller and Rubinow \cite{keller1976slender}) Stokes drag force $\bm{F}$ on a cylinder of length $2\sqrt{1/4+\epsilon^2}$ with hemispherical endpoints and with centerline along the $x$-axis. Columns 2 and 3 compare the $y$-component of $\bm{F}$ for a cylinder translating with unit speed in the $y$-direction, while columns 4 and 5 compare the $x$-component of $\bm{F}$ for translation in the $x$-direction. Here the expected $\epsilon\log\epsilon$ scaling of the difference between forces is less apparent, particularly in the $y$-direction. This may be due to endpoint effects (see figure \ref{fig:cylinder}). }\label{validate3}
\end{table}

\begin{figure}[!h]
	\centering
	\includegraphics[scale=0.35]{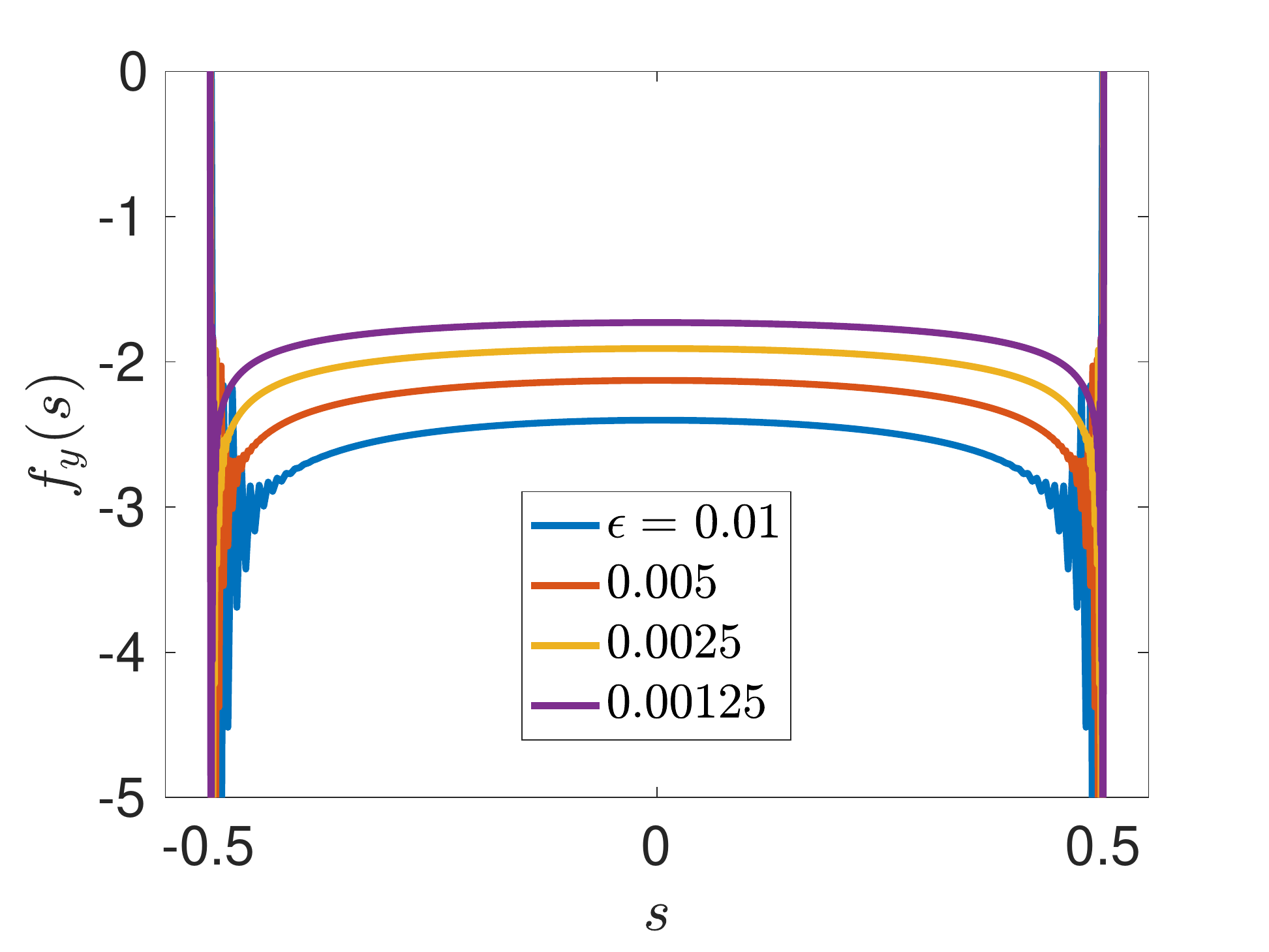}
	\includegraphics[scale=0.35]{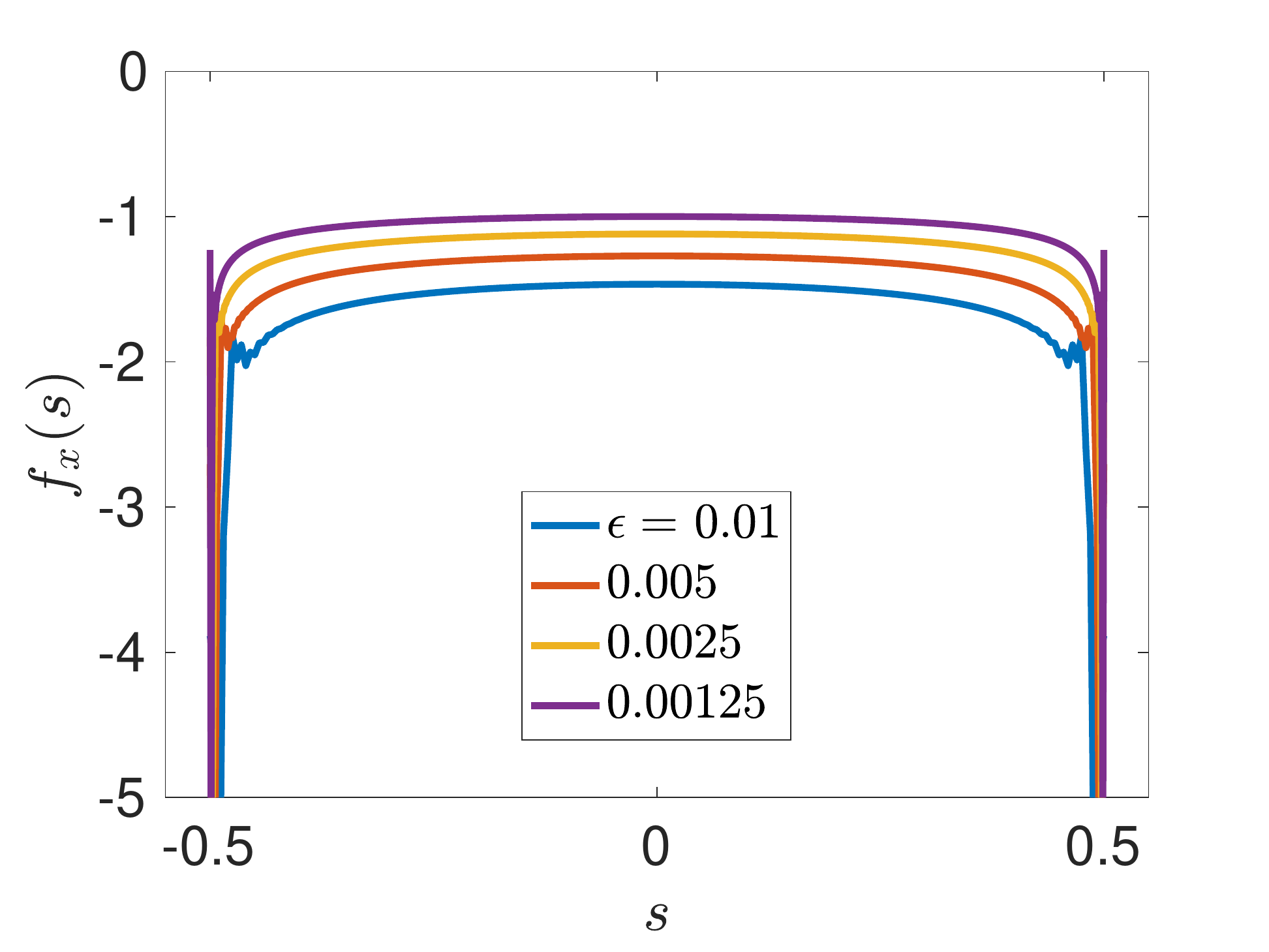}
	\caption{Force per unit length $\bm{f}(s)$, $s\in [-1/2,1/2]$, along the uniform cylinder with hemispherical caps at the endpoints and centerline aligned with the $x$-axis. The left figure shows the $y$-component of the force density for the cylinder translating with unit speed in the $y$-direction, while the right figure shows the $x$-component of the force density for the cylinder translating in the $x$-direction. Comparing with figure \ref{fig:prolate}, it is clear that the shape of the radius function $r(s)$ at the fiber endpoint has a large effect on $\bm{f}(s)$. In particular, despite the decay in $\bm{f}(s)$ at the very endpoint of the fiber, the oscillations leading up to the endpoint brings the physical validity of this force density into question. }
	\label{fig:cylinder}
\end{figure}

The computed drag force in table \ref{validate3} agrees well with the asymptotic expression of Keller and Rubinow \cite{keller1976slender}; however, the computed force-per-unit-length $\bm{f}(s)$ is not as physically reasonable at the fiber endpoints. According to \cite{free_ends}, in the case of a cylinder with hemispherical caps, we actually want a faster rate of decay in the force near the fiber endpoints -- in particular, we need $\bm{f}(s)/(1/4-s^2)\in C(-1/2,1/2)$. However, as shown in figure \ref{fig:cylinder}, flow about the cylinder results in wild oscillations in $\bm{f}(s)$ near the fiber endpoints. Possibly this indicates that this method (and likely others based on slender body theory) are really designed to treat prolate spheroids with sufficient decay in radius near the fiber endpoints.  

\subsection{Closed loops in constant flow}\label{loop_valid}
To validate the slender body approximation \eqref{SB_new} in the closed loop setting ($\mc{I}=\T$), we compute the Stokes drag about a translating thin torus of length 1 with centerline in the $xy$-plane and axis of symmetry about the $z$-axis. We compare the computed drag force for various values of $\epsilon$ to the asymptotic expression of Johnson and Wu \cite{johnson1979hydromechanics}  (see table \ref{tab:validate1}). Note that for the thin filaments that we consider here, the Johnson and Wu expression for the drag force corresponds well with the semianalytic expression for a torus translating in the $z$-direction, derived by Majumdar and O'Neill \cite{majumdar1977axisymmetric} with corrections by Amarakoon, et al. \cite{amarakoon1982drag}. The Majumdar-O'Neill expression, consisting of an infinite sum of Legendre functions, holds for general values of $s_0$, where $s_0$ is defined to be the ratio of the outer radius of the torus (measured centerline to longitudinal axis) to the cross sectional radius. In \cite{amarakoon1982drag}, Amarakoon, et al. numerically verify the reported $\mc{O}(s_0^{-2})$ accuracy of the Johnson-Wu expression. In our case, we are mainly concerned with the parameter region $s_0 = 1/(2\pi\epsilon)>10$, so the Johnson-Wu expression agrees with the exact expression for Stokes drag in the $z$-direction to at least two digits. 

\begin{figure}[!h]
	\centering
	\includegraphics[scale=0.35]{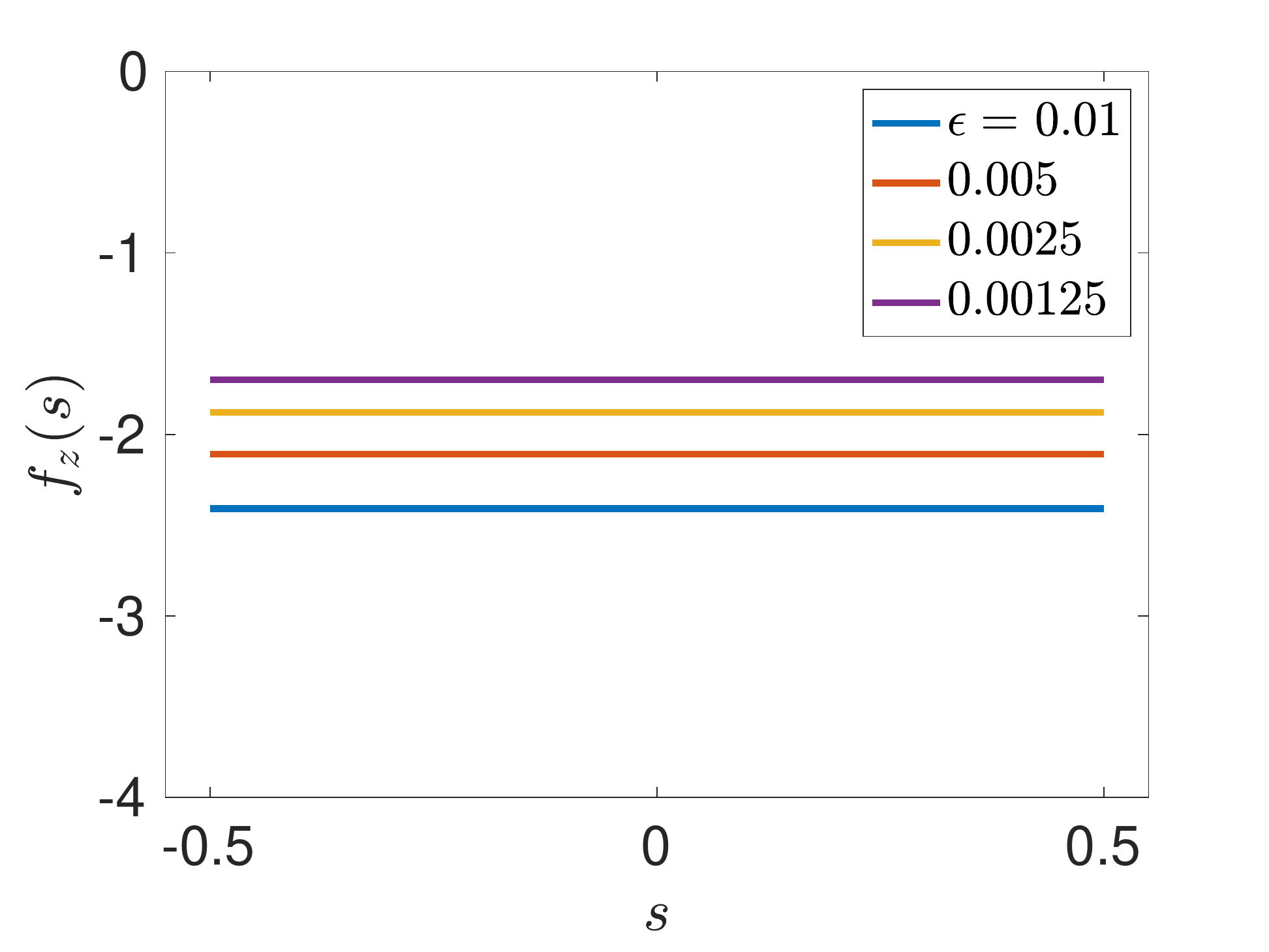}
	\includegraphics[scale=0.35]{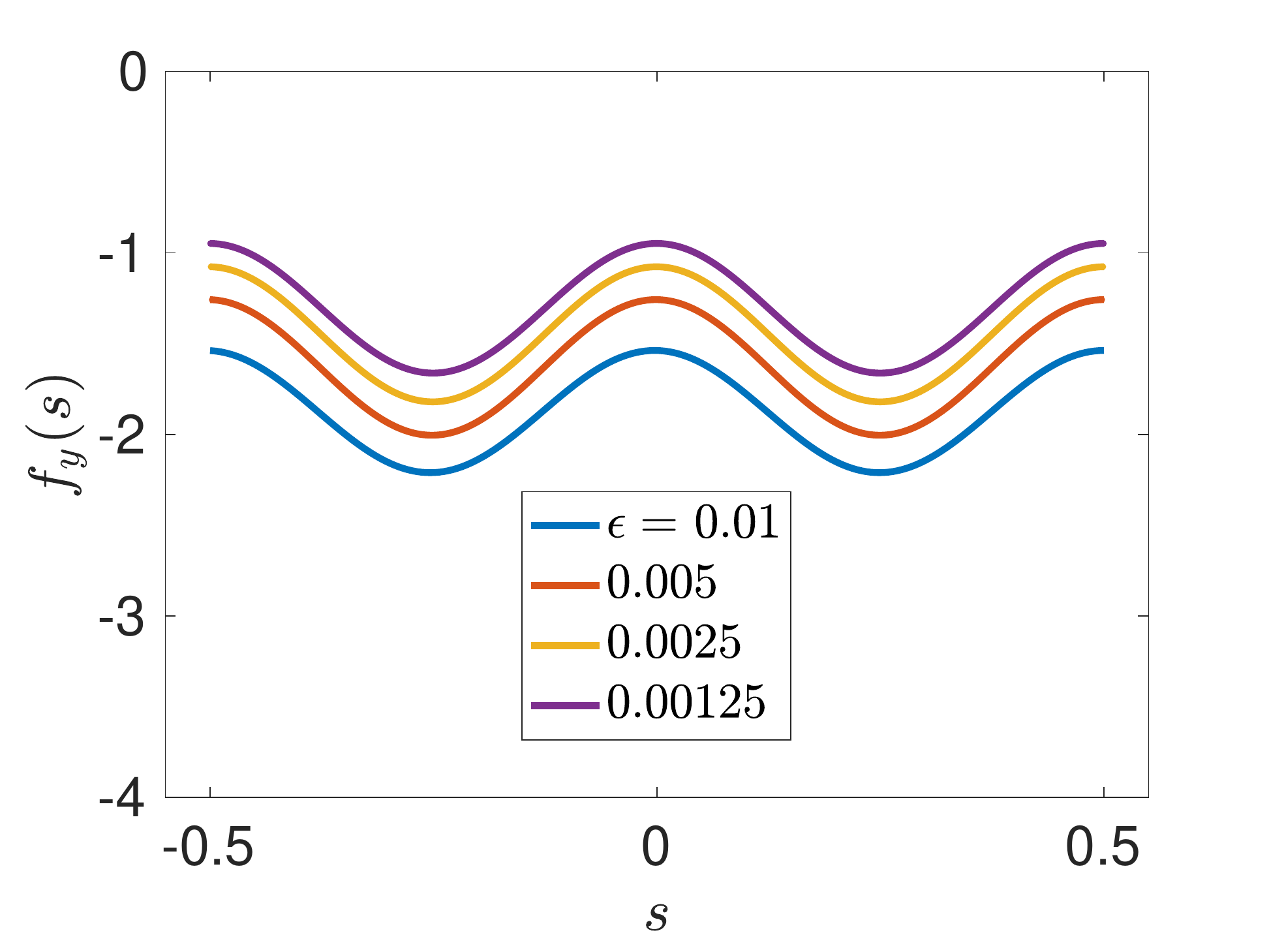}
	\caption{Force per unit length $\bm{f}(s)$, $s\in \T$, along the slender torus with centerline in the $xy$-plane. The left figure shows the $z$-component $\bm{f}(s)\cdot\be_z$ for a slender body translating with unit speed in the $z$-direction, while the right picture shows the $y$ component $\bm{f}(s)\cdot\be_y$ for translation with unit speed in the $y$-direction.   }
	\label{fig:ring}
\end{figure}

Since the torus centerline $\X(s)$ is planar, we choose the normal vector $\cos(2\pi s) \be_x + \sin(2\pi s)\be_y$ to also lie in the $xy$-plane. The integral term in \eqref{SB_new} is discretized using the trapezoidal rule, and the number of discretization points $n$ along the fiber centerline is taken to be $n=2/\epsilon$. Given zero background flow and uniform unit speed in the $z$-direction (columns 2 and 3, table \ref{tab:validate1}) and $y$-direction (columns 4 and 5, table \ref{tab:validate1}), the discretized operator \eqref{SB_new} is inverted to find the force per unit length $\bm{f}(s)$, which is then summed over $s$ to find the drag force $\bm{F}$. We plot the calculated $\bm{f}(s)$ in figure \ref{fig:ring} to verify that the computed force density makes physical sense. For all computations, we take the viscosity $\mu=1$.  

\begin{table}[!h]
	\centering
	\begin{tabular}{ | c | c | c || c | c | c |} 
		\hline
		& \multicolumn{2}{|c||}{ $\bm{F}\cdot\be_z$ for $\bu=\be_z$} & \multicolumn{2}{|c|}{$\bm{F}\cdot\be_y$ for $\bu=\be_y$} & \\
		\hline
		$\epsilon$  & Eqn \eqref{SB_new} & Johnson-Wu & Eqn \eqref{SB_new} & Johnson-Wu  & $\epsilon\abs{\log\epsilon}$ \\ 
		\hline
		0.01 & -2.4093 & -2.3503 & -1.8740 & -1.8292 & 0.0461 \\ 
		0.005 & -2.1076 & -2.0806 & -1.6309 & -1.6103 & 0.0265 \\ 
		0.0025 & -1.8788 & -1.8664 & -1.4484 & -1.4389 & 0.0150 \\ 
		0.00125 & -1.6979 & -1.6922 &  -1.3051 & -1.3007 & 0.0084 \\
		\hline
	\end{tabular}
	\caption{We consider a translating slender torus of length 1 with centerline lying in the $xy$-plane, and compare the resulting Stokes drag force given by the slender body model (expression \eqref{SB_new}) to the asymptotic expression calculated by Johnson and Wu \cite{johnson1979hydromechanics}. Columns 2 and 3 compare the $z$-component of the drag force for a slender torus translating with speed 1 in the $z$-direction (``broadwise translation''), while columns 4 and 5 show the $y$-component of the drag for translation in the $y$-direction (``translation perpendicular to the longitudinal axis''). Here we can see an approximate $\epsilon\log\epsilon$ scaling in the difference between the two expressions. } \label{tab:validate1}
\end{table}

Our method agrees quite well with the asymptotic expression of Johnson and Wu -- as expected, table \ref{tab:validate1} shows roughly an $\mc{O}(\epsilon\log\epsilon)$ difference between the slender body approximation to the drag force and the asymptotic expression. This is encouraging since both \eqref{SB_new} and the Johnson-Wu asymptotics are based on the Stokeslet/doublet expression \eqref{stokes_SB}. We have chosen these particular values of $\epsilon$ so that our method can also be compared with the regularized Stokeslet method of Cortez and Nicholas \cite{cortez2012slender}. 

\subsection{Free ended fibers in shear flow}

\begin{figure}
	\centering
		\begin{subfigure}{0.45\textwidth}
		\includegraphics[width=\linewidth]{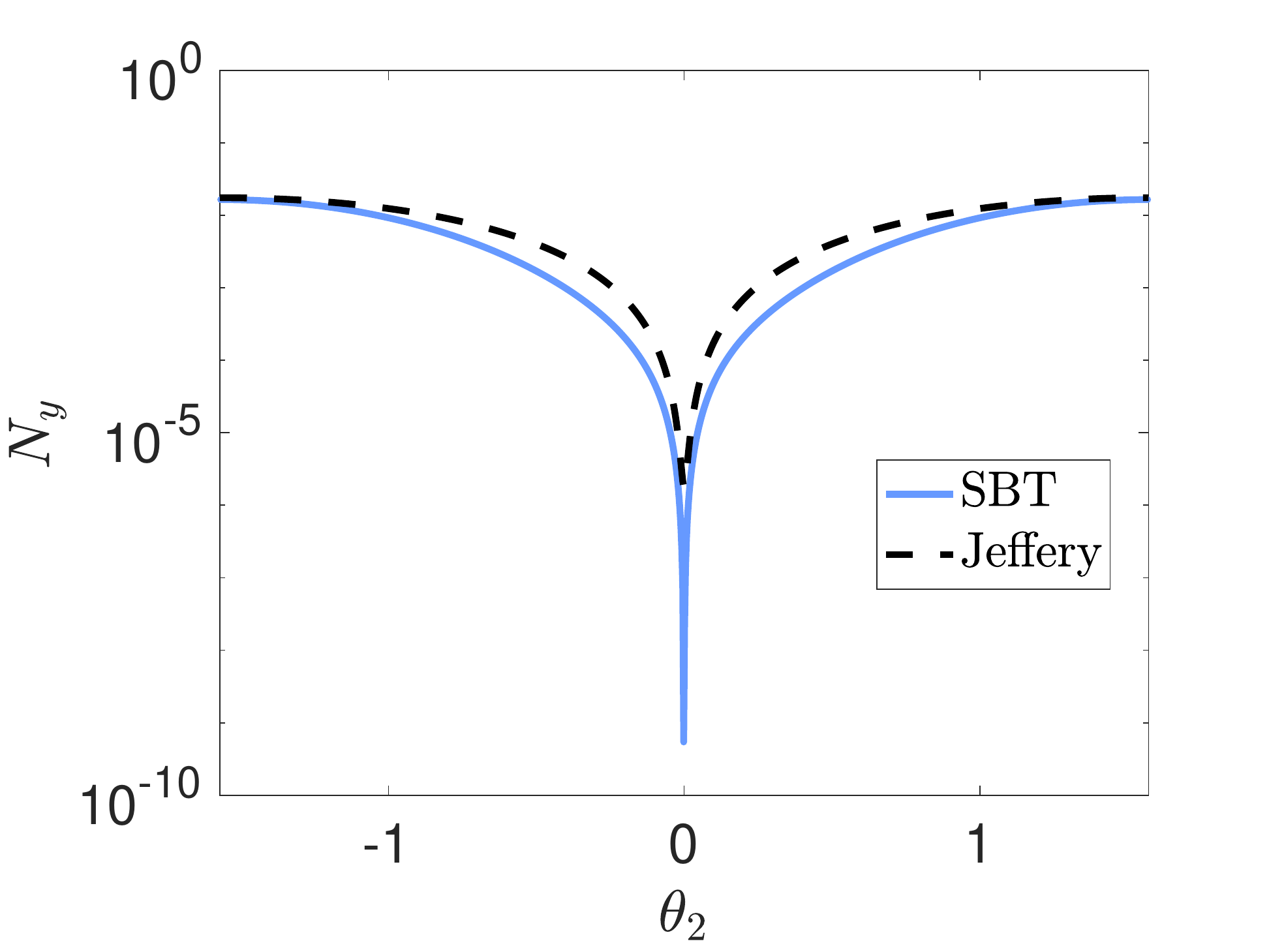}
		\caption{}
		\label{fig:nvsthetalambda100n200}
	\end{subfigure}
	\begin{subfigure}{0.45\textwidth}
		\includegraphics[width=\linewidth]{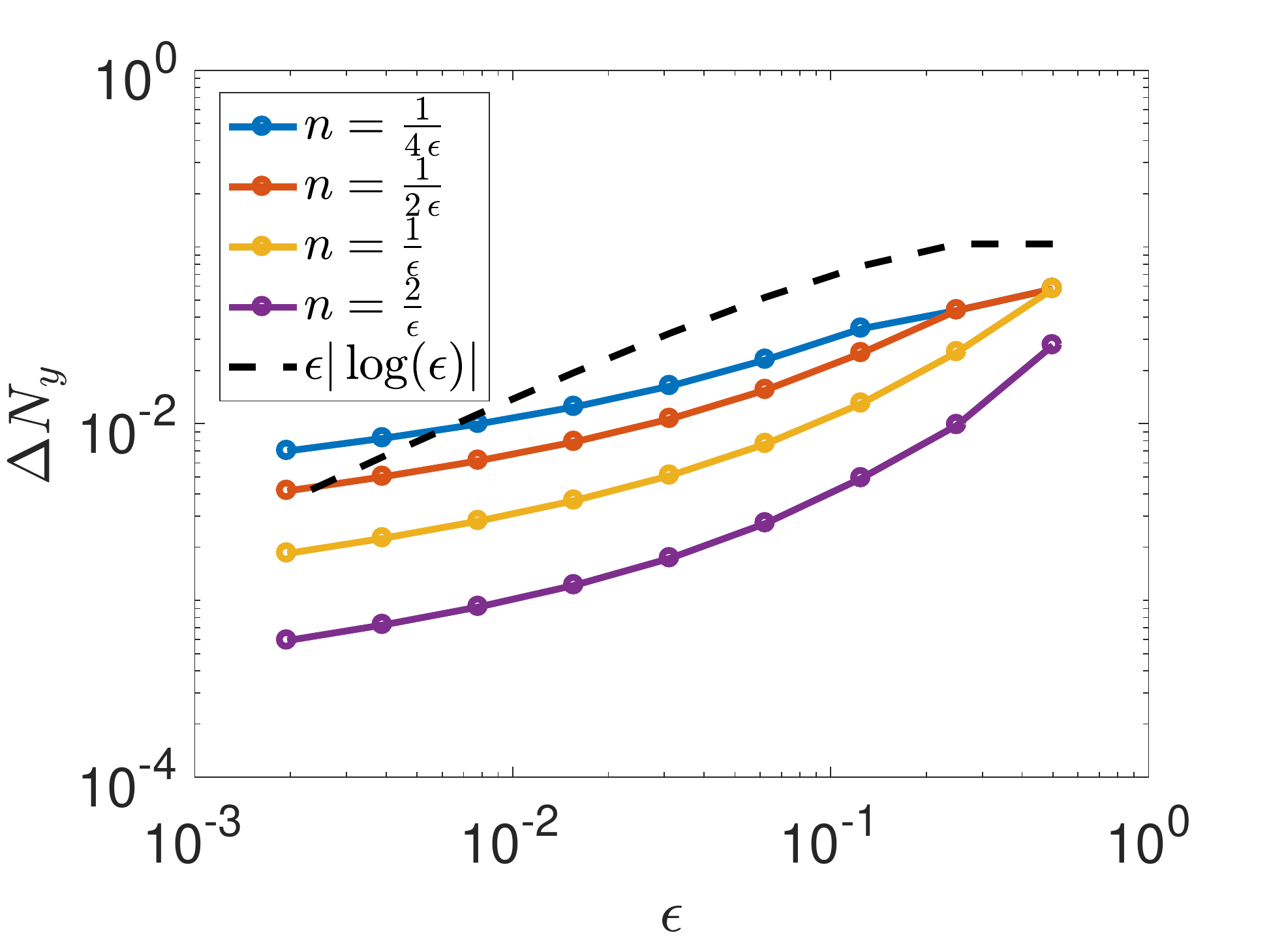}
		\caption{}
		\label{fig:FIG_dN_vs_eps}
	\end{subfigure}
	\caption{(a) The $y$-component of the torque for a prolate spheroid with $\lambda = 100$ for different orientations in shear flow. The values $\theta_2=0,\pm \pi/2$ correspond to alignment parallel and perpendicular to the shear plane, respectively.  (b) The difference $\Delta N_y$ between the $y$-component of the torques due to Jeffery and slender body theory for a prolate spheroid of aspect ratio $\lambda = 1/\epsilon$ aligned in the $z$-direction in shear flow. }
\end{figure}

In this section we calculate the angular momentum of a prolate spheroid with aspect ratio $\lambda=1/\epsilon$ in the shear flow field $\bu(z) = (z,0,0)^{\rm T}$. The torques are derived using both slender body theory (equation \eqref{SB_new}) and the Jeffery model (equation \eqref{Jeffery}) for comparison. Figure \ref{fig:nvsthetalambda100n200} shows how the torque of the ellipsoid varies as a function of its orientation. Here, $\theta_2$ is the second Euler angle and $\theta_2=[-\pi/2,\pi/2]$ corresponds to a full revolution about the $y$-axis. We see that the torques agree at $\theta_2=\pm\pi/2$ and the discrepancy between the two models increases as the orientation approaches alignment in the shear plane; in particular, the torque in the slender body model goes to zero but the Jeffery torque remains bounded away from zero. Since the fluid velocity is exactly zero along the particle centerline, the slender body model does not yield a torque on the particle. On the other hand, in the Jeffery model, the spheroid is aware of the non-zero fluid velocity gradient, and hence experiences a non-zero torque at this orientation. 

Figure \ref{fig:FIG_dN_vs_eps} shows the difference between the $y$-component of the torques due to Jeffery and slender body theory as a function of $\epsilon$ for different values of $n$.  The particles are oriented with $\theta_2=\pi/2$, perpendicular to the shear plane. We see roughly $\mathcal{O}(\epsilon\log(\epsilon))$ convergence for the five largest values of $\epsilon$. For smaller values of $\epsilon$, the model converges at a slower rate. This is similar to the observed convergence in the force values (table \ref{validate2}), which are calculated for $\epsilon \le10^{-2}$. In addition, the two models show better agreement as the discretization parameter $n$ is increased. 

\begin{figure}
	\centering
	\begin{subfigure}{0.45\textwidth}
		\includegraphics[width=\linewidth]{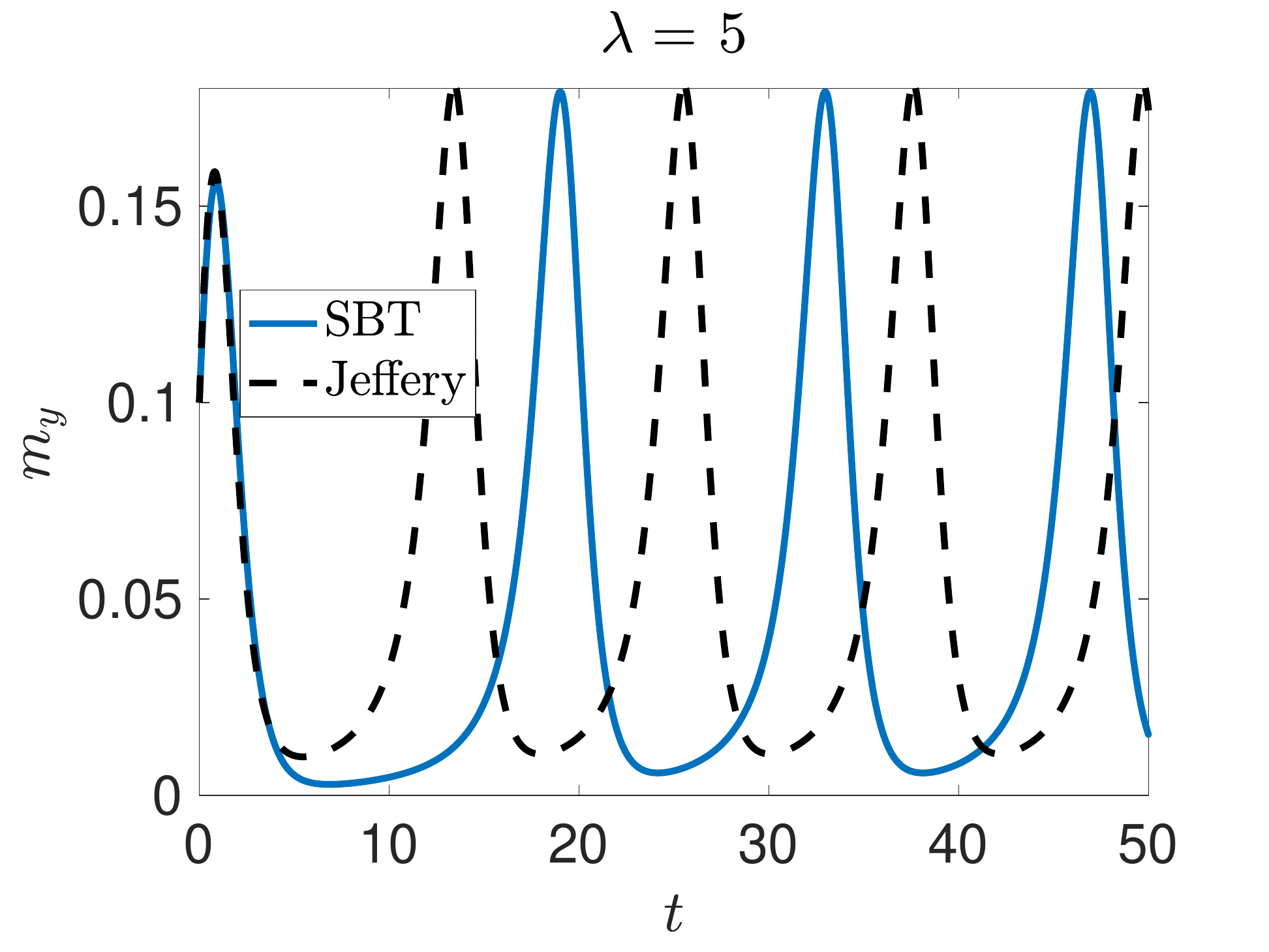}
		\caption{}
		\label{fig:fignyvstlam5}
	\end{subfigure}
	\begin{subfigure}{0.45\textwidth}
		\includegraphics[width=\linewidth]{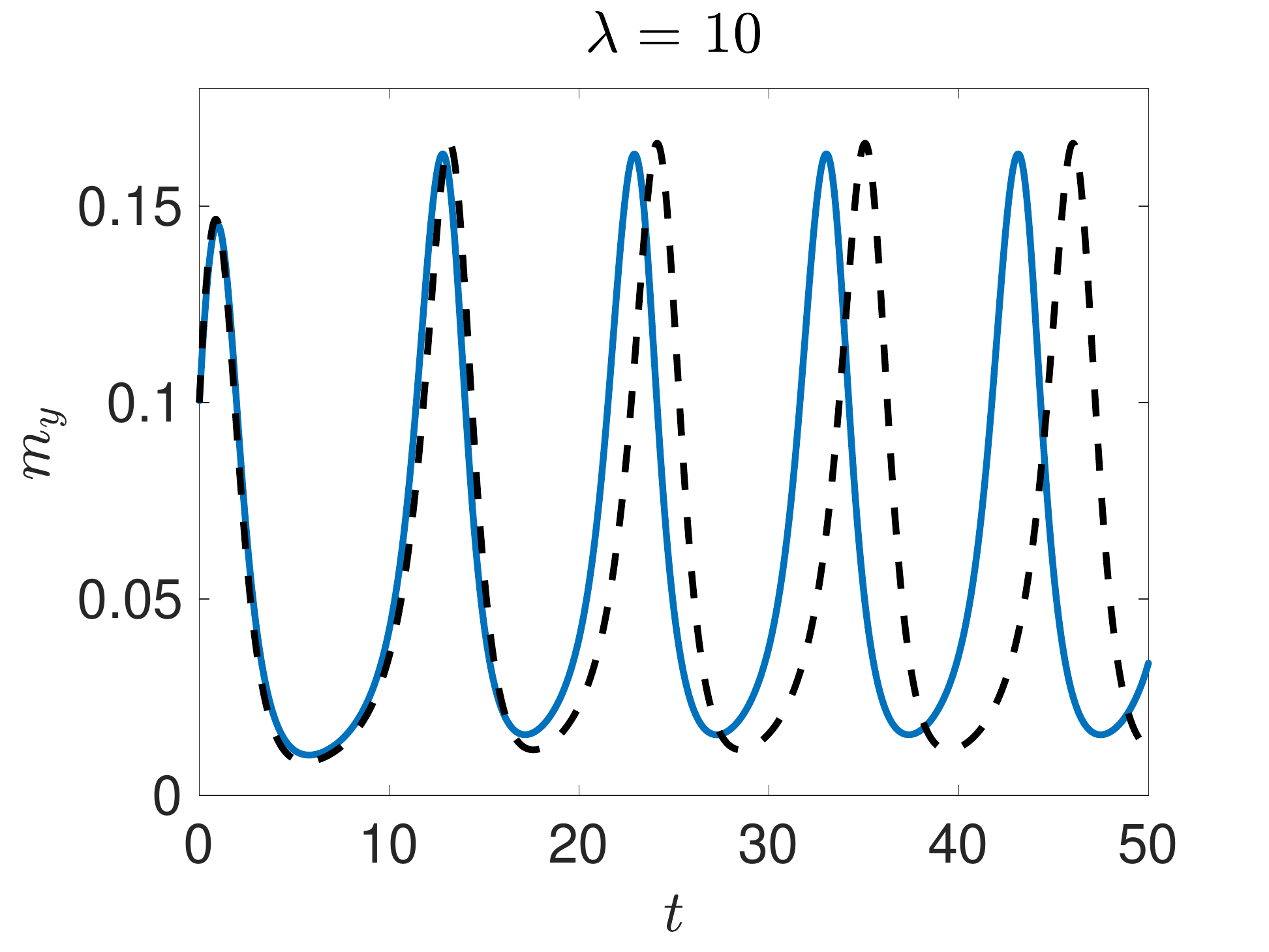}
		\caption{}
		\label{fig:fignyvstlam10}
	\end{subfigure}
	
	\begin{subfigure}{0.45\textwidth}
		\includegraphics[width=\linewidth]{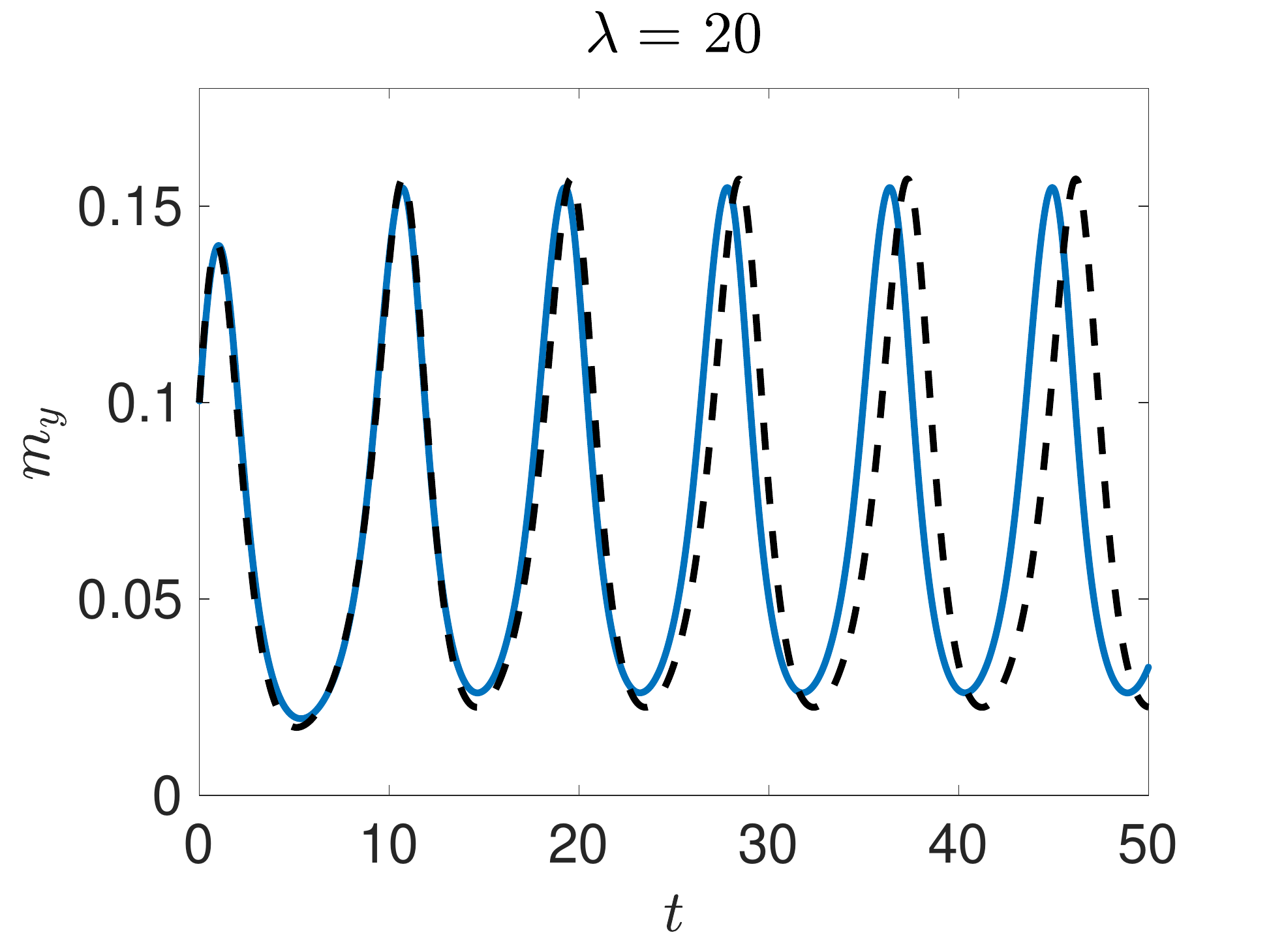}
		\caption{}
		\label{fig:fignyvstlam20}
	\end{subfigure}
	\begin{subfigure}{0.45\textwidth}
		\includegraphics[width=\linewidth]{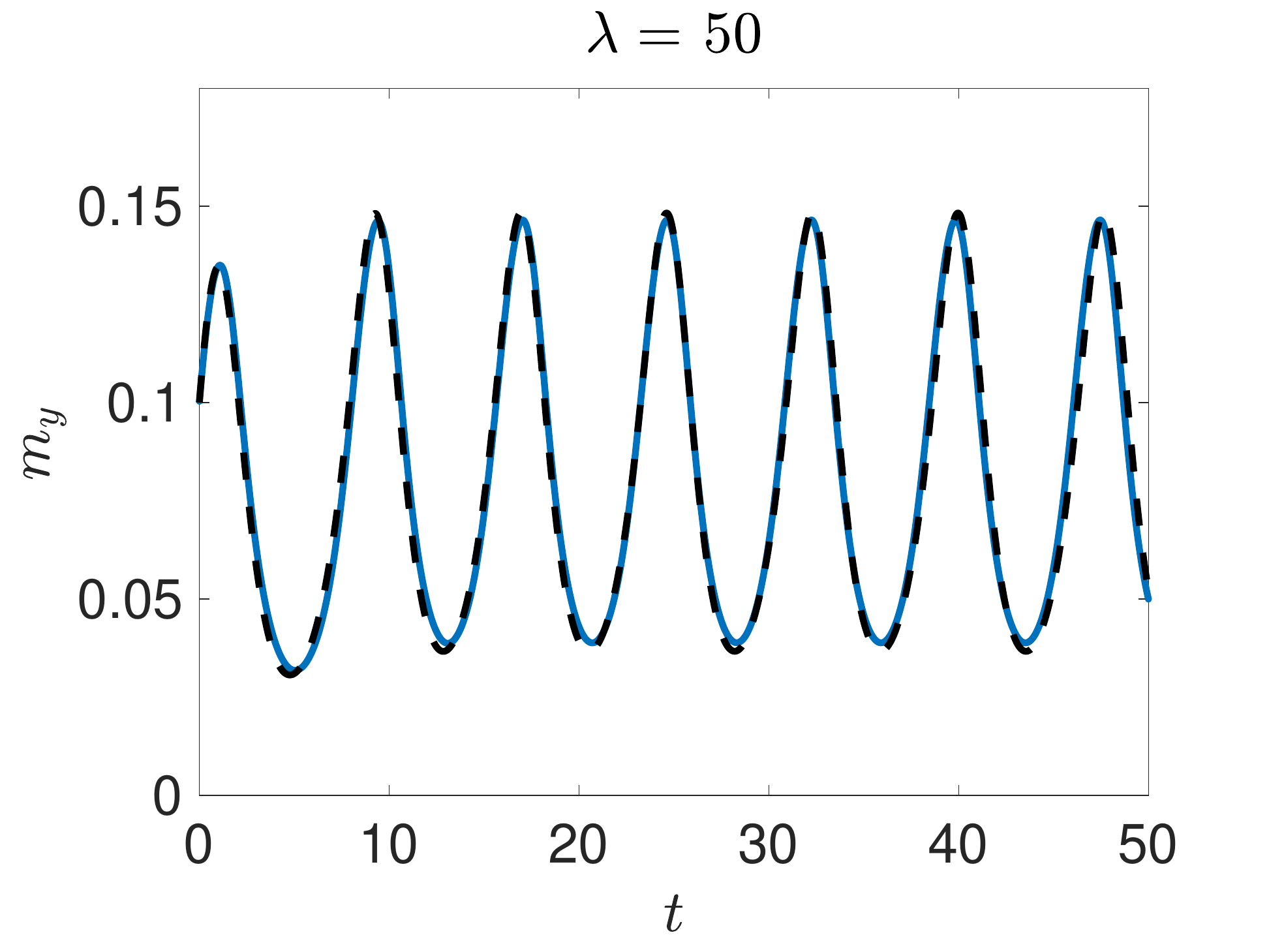}
		\caption{}
		\label{fig:fignyvstlam50}
	\end{subfigure}

	\begin{subfigure}{0.45\textwidth}
	\includegraphics[width=\linewidth]{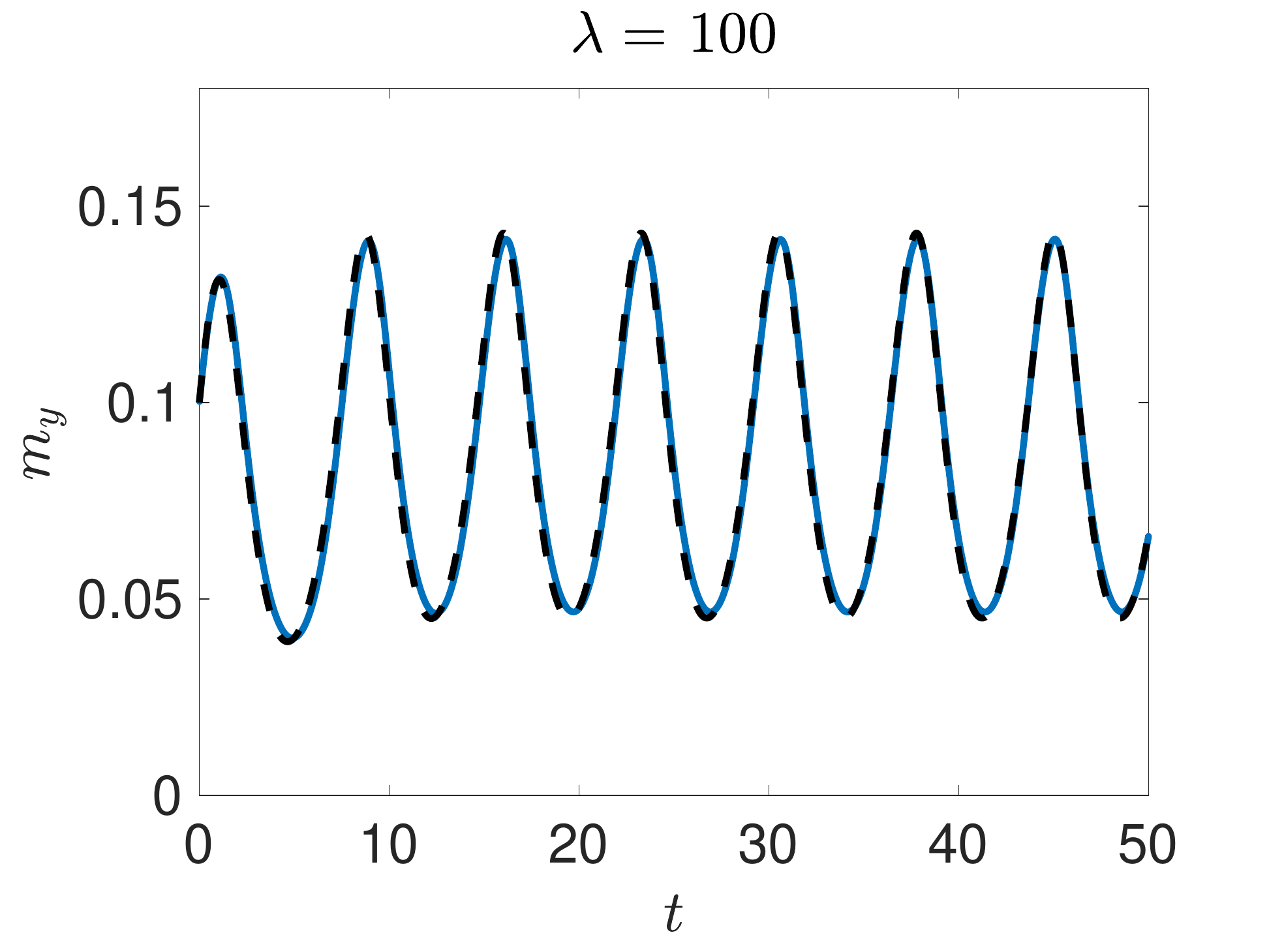}
	\caption{}
	\label{fig:fignyvstlam100}
\end{subfigure}
	\caption{The $y$-component of the angular momentum of a particle in shear flow calculated from slender body theory (blue) and Jeffery (black, dashed). The aspect ratio takes different values in the range $\lambda\in[10,100]$. The simulation parameters are $\mu = 0.06$, $n = \frac{2}{\epsilon}$}\label{dynamics_lam}
	%
	
\end{figure}

Figure \ref{dynamics_lam} shows the the $y$-component from equation \eqref{eq:rotation} of the torques due to slender body theory and Jeffery. The ODE for angular momentum is solved using one of MATLAB's built in functions such as \texttt{ode15s}. The particles are aligned as before with initial conditions $\bm{m}_0 = (0,0.1,0)^{\rm T}$ and Euler angles $(0,\pi/2,0)^{\rm T}$; hence the only non-zero component of the angular momentum is $m_y$. We observe that for a relatively low aspect ratio (i.e., figure \ref{fig:fignyvstlam5}) the models do not agree so well, however $\lambda = 5$ is not considered to be in the ``slender" regime and we therefore do not expect good agreement here. As $\lambda$ increases, the dynamics become almost indistinguishable. 

\begin{figure}
	\centering
		\begin{subfigure}{0.45\textwidth}
		\includegraphics[width=\linewidth]{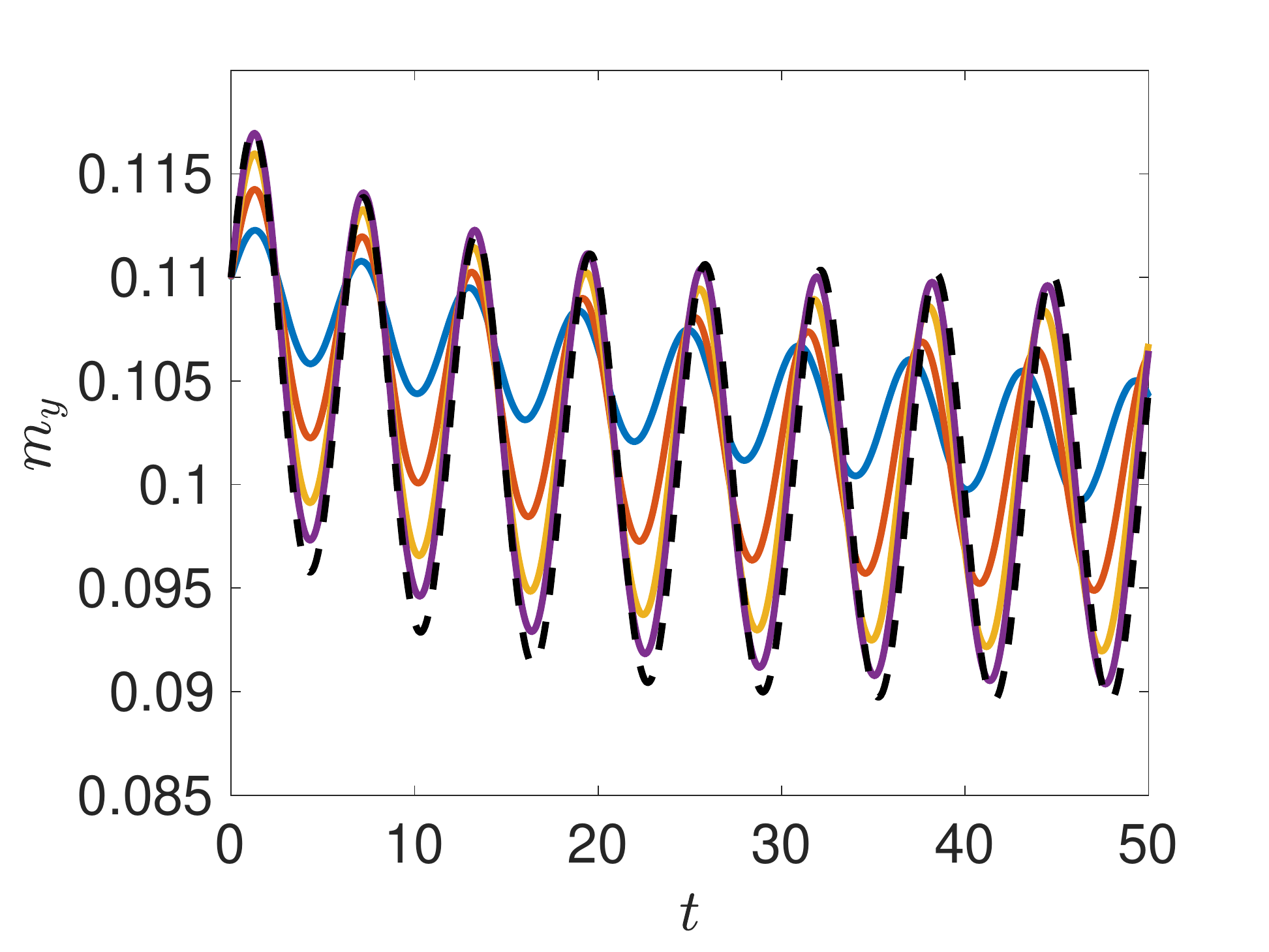}
		\caption{}
		\label{fig:figmtvaryingn}
	\end{subfigure}
	\begin{subfigure}{0.45\textwidth}
	\includegraphics[width=\linewidth]{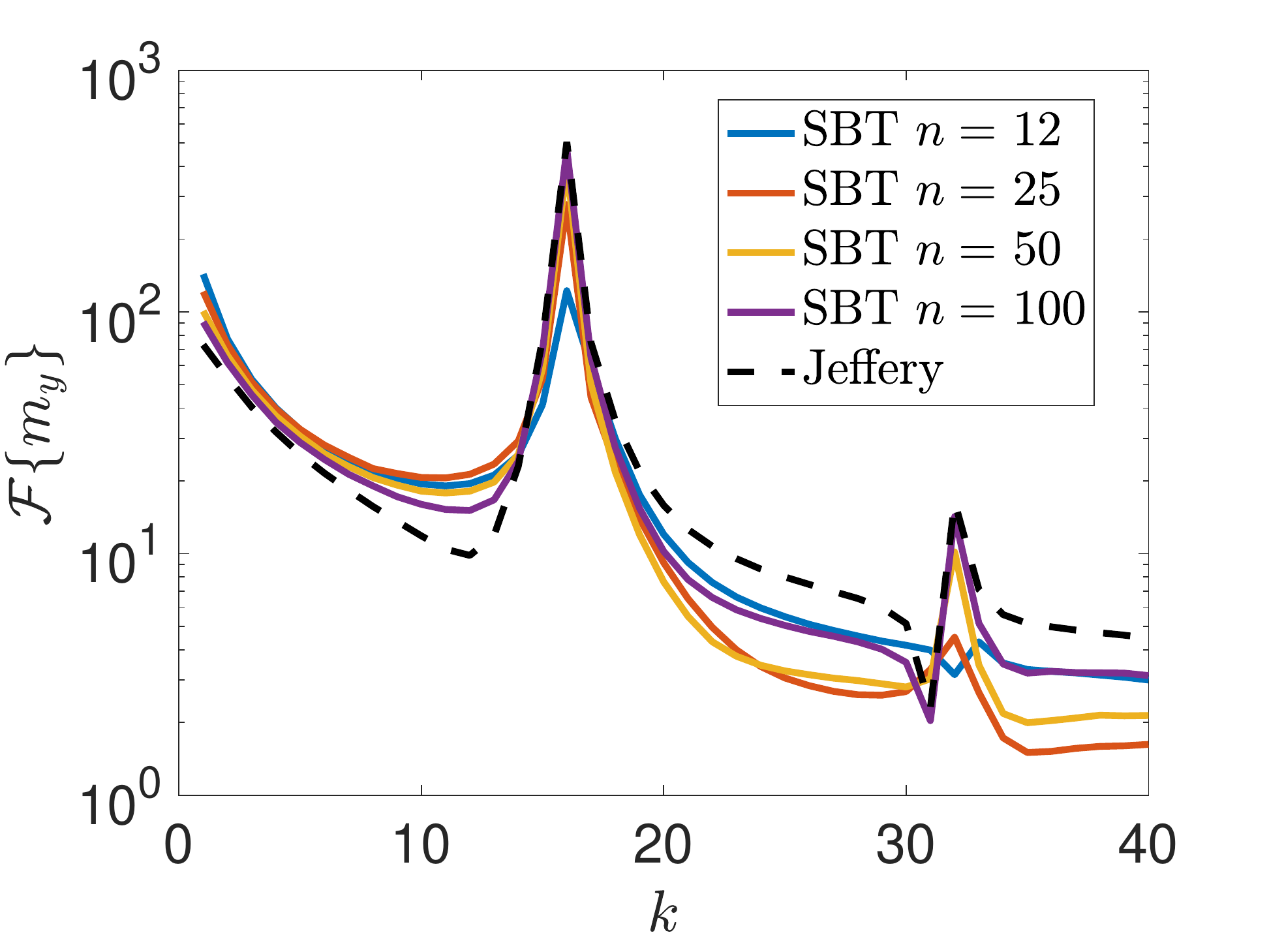}
\caption{}
\label{fig:figfouriermtvaryingn}
	\end{subfigure}
\caption{The $y$-component of the angular momentum of a particle in shear flow (a) and the first 40 Fourier modes (b). The colored lines are calculated from slender body theory with discretization parameter varying in the range $n\in[12,100]$ and the dashed line is due to Jeffery. The simulation parameters are $\mu = 0.01$ and $\lambda = 50$ and $\bm{m}_0 = (0,0.11,0)^{\rm T}$.}\label{dynamics_n}
%
%

\end{figure}

We now turn our attention to figure \ref{dynamics_n}, which displays how the choice of the discretization parameter $n$ affects the solution quality. Figure \ref{fig:figmtvaryingn} shows $m_y$ for the slender body model for different numbers of discretization points $n$ and figure \ref{fig:figfouriermtvaryingn} shows its 40 highest Fourier modes. The main observation here is that the model becomes more accurate as $n$ increases. In particular, if $n$ is chosen to be too low (here, too low corresponds to roughly less than $1/(2 \epsilon)$) then the model does not resolve the low frequency modes, which can be seen by the spike at $k=16$ in figure \ref{fig:figfouriermtvaryingn}, where only the $n=50$ and $100$ lines are able to reasonably capture this mode correctly.

\subsection{Closed loops and oblate spheroids in shear flow}
In this section we compare the rotational dynamics of a thin torus modeled by slender body theory to the rotational dynamics of an oblate disk of similar shape and mass. This comparison differs from the prolate spheroid comparisons in that here the particle shapes are different and we do not expect the two solutions to coincide. The slender torus experiences a force only along its centerline, whilst the oblate spheroid experiences a force all across its surface. In addition, the moment of inertia tensor for a torus of inner radius $2\epsilon$ and of outer radius $a$ (measured from the center of mass to the centerline) is given by
\begin{equation}
	J_T = m_T\,\mathrm{diag}\left( \frac{ 4\,a^2+5\,\epsilon ^2}{8}, \frac{ 4\,a^2+5\,\epsilon ^2}{8} ,\frac{4\,a^2+3\,\epsilon ^2}{4}\right).
\end{equation}
Setting the mass of the torus to $m_T = 2\,m_p/5$, where $m_p$ is the mass of the spheroid, we have the relation $J - J_T = \mathcal{O}(\epsilon^2)$ for an oblate spheroid with semi minor axis length $b = \epsilon$. Due to the particle shape, the oblate spheroid experiences a much stronger torque; hence for the torques to be of the same magnitude, a viscosity of $\mu_T = 200 \mu$ is chosen for the torus. The particles are placed at rest in the shear flow with the initial Euler angles $(0.01,0.01,0.01)$. We do this for two reasons: the first being that the Euler angles $(0,0,0)$ correspond to a neutrally stable orbit where the ellipsoid exhibits a tumbling motion forever. The second reason is that these angles correspond to exact alignment in the $xy$ plane, where the slender model will not experience a force since the fluid velocity is exactly zero.

\begin{figure}
	\begin{subfigure}{0.45\textwidth}
	\includegraphics[width=\linewidth]{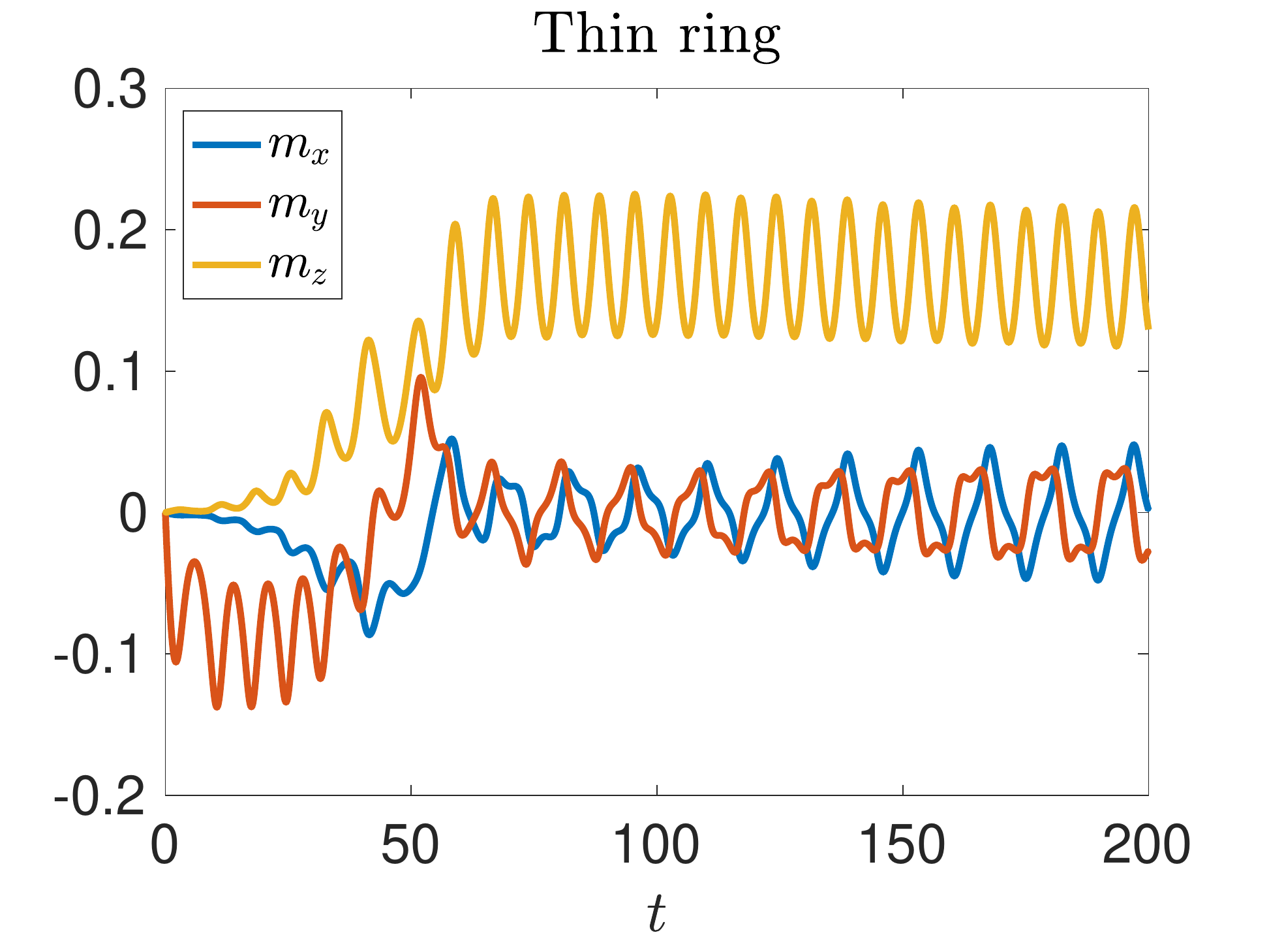}
	\caption{}
	\label{fig:ring1}
\end{subfigure}
	\begin{subfigure}{0.45\textwidth}
		\includegraphics[width=\linewidth]{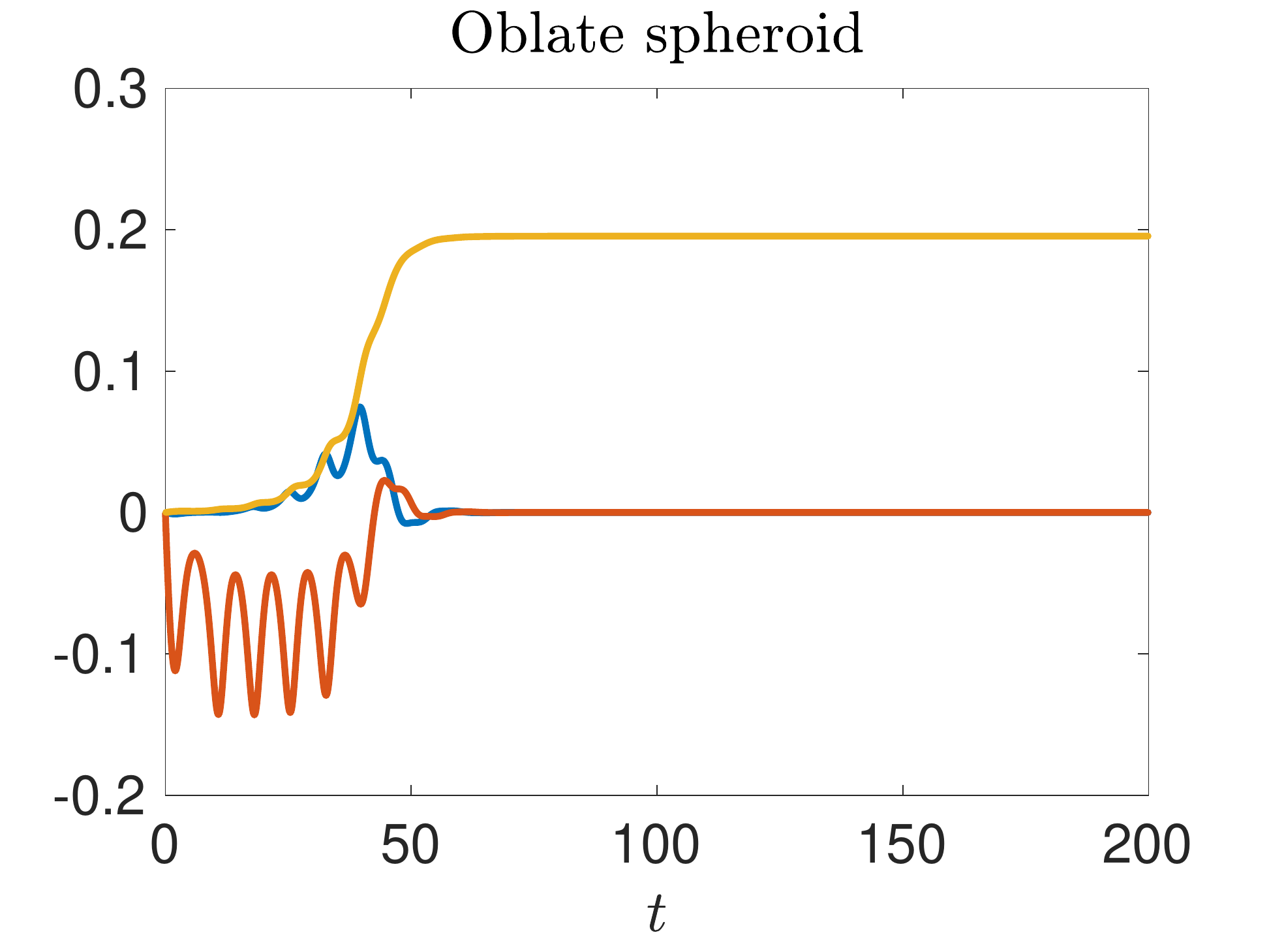}
		\caption{}
		\label{fig:oblate1}
	\end{subfigure}

	\begin{subfigure}{0.45\textwidth}
	\includegraphics[width=\linewidth]{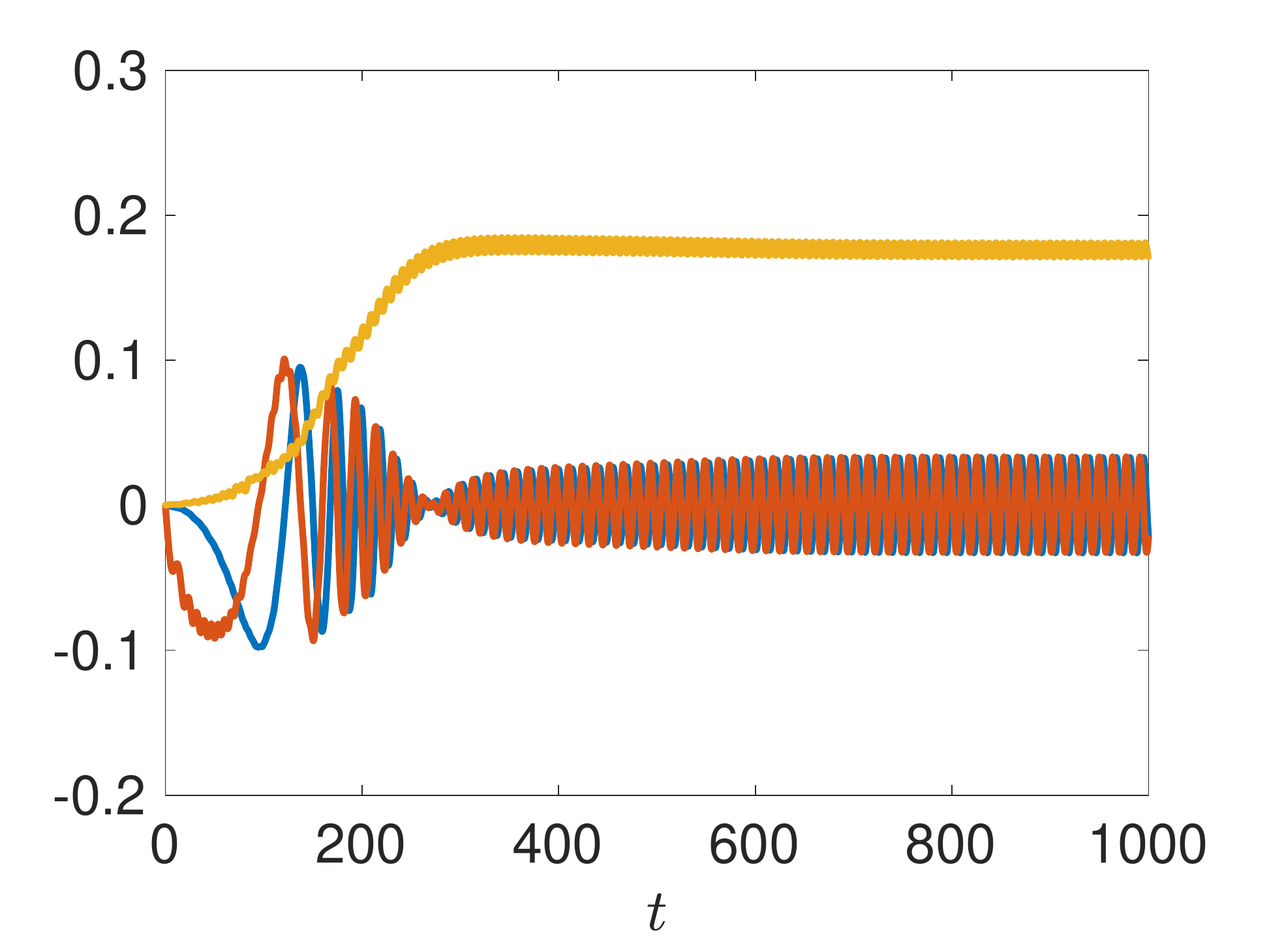}
	\caption{}
	\label{fig:ring2}
\end{subfigure}
\begin{subfigure}{0.45\textwidth}
	\includegraphics[width=\linewidth]{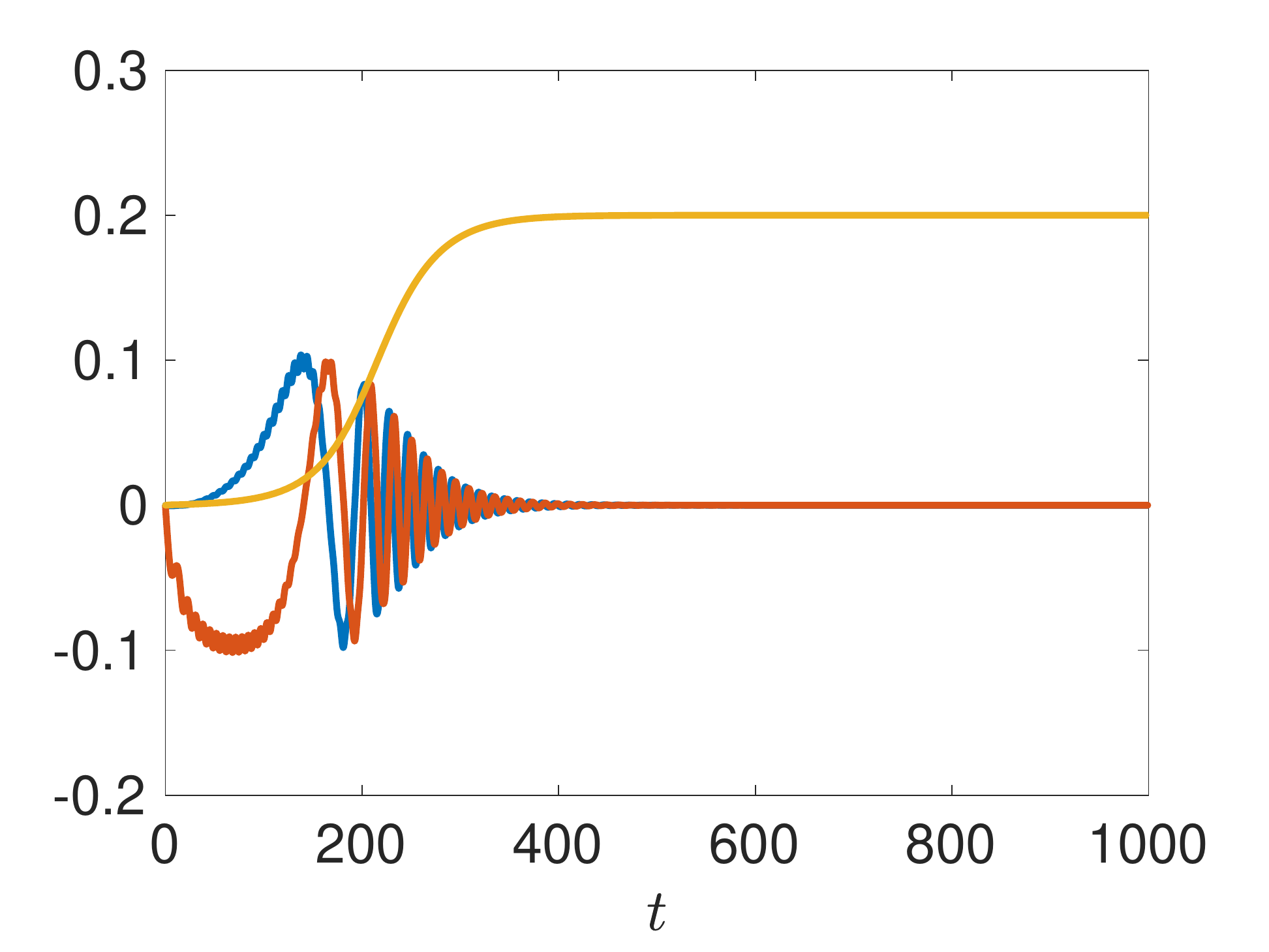}
	\caption{}
	\label{fig:oblate2}
\end{subfigure}
	
	\begin{subfigure}{0.45\textwidth}
	\includegraphics[width=\linewidth]{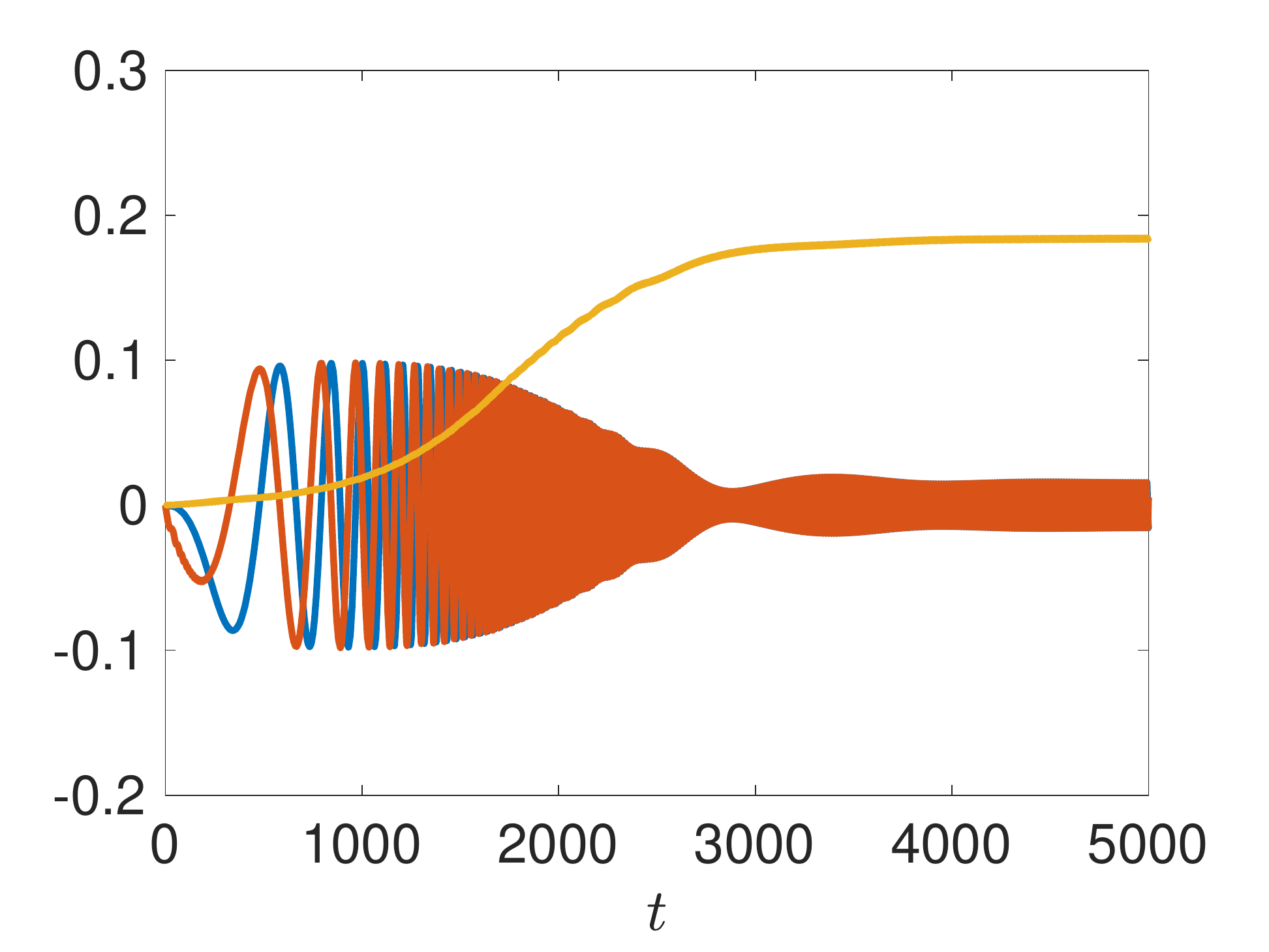}
	\caption{}
	\label{fig:ring3}
\end{subfigure}
\begin{subfigure}{0.45\textwidth}
	\includegraphics[width=\linewidth]{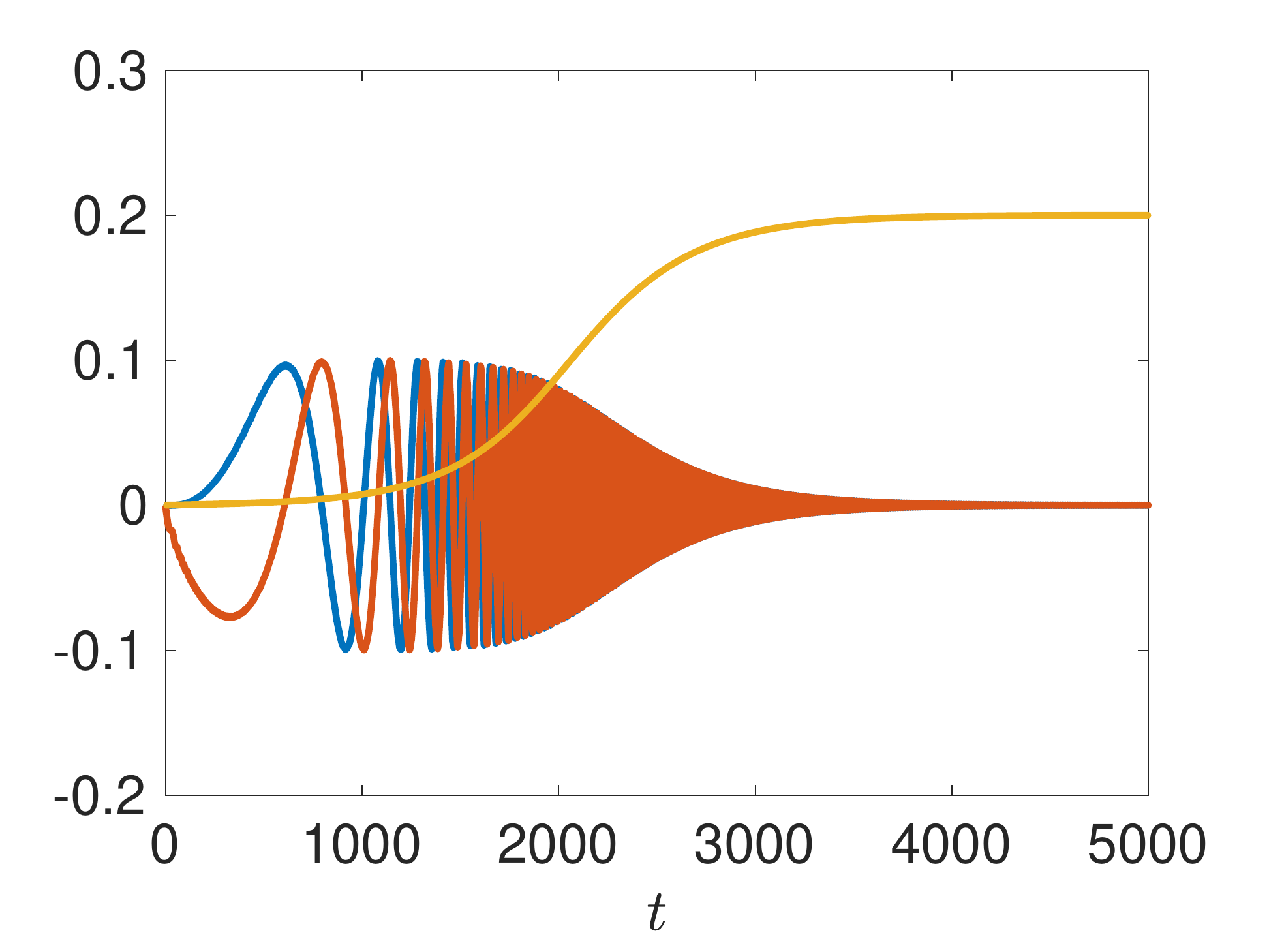}
	\caption{}
	\label{fig:oblate3}
\end{subfigure}
	\caption{The angular momentum components of a thin ring (left column) and an oblate spheroid (right column) for $\mu_0=0.01$, $0.001$ and $0.0001$ (from top to bottom). The particle parameters are $\epsilon = \frac{1}{100}$, $\lambda = \frac{1}{100}$, $\mu_{T} = 200\mu$, $m_{T} = \frac{2}{5}m$, $m=1$, $a=1$, $n=\frac{1}{2 \epsilon}$}\label{dynamics_ring}
	%
	
\end{figure}

 Challabotla et al. \cite{challabotla2015rotational} conduct a similar experiment with oblate spheroids in shear flow and observe two phases of rotation: (1) an unstable wobbling phase of length proportional to the particle inertia, and (2) a stable rolling phase, where the spheroid aligns and rolls perfectly in the shear plane. Figure \ref{dynamics_ring} shows $\bm{m}(t)$ for the thin ring with $\epsilon=1/100$ and oblate spheroid with $\lambda=1/100$ for three different values of $\mu$ (and the corresponding values of $\mu_T$). For the spheroid model, we observe the temporary initial wobbling phase followed by the stable rolling phase where the particle rotates in the shear plane with a constant $m_z$ component. In addition, as the relative particle inertia increases (that is, as the $\mu$ decreases), the wobbling phase is prolonged. These two observations are in agreement with the results in \cite{challabotla2015rotational}. If we turn our attention to the thin ring, we observe some similarities: there is an initial wobbling phase followed by a somewhat different rolling phase. In the rolling phase, the particle's symmetry axis (the $z$-axis in the particle frame) precesses about the $y$-axis in the inertial frame. This is seen as oscillations in the $m_x$ and $m_y$ components about a mean zero value, which in turn affects the $m_z$ component. A possible explanation for this precession is the fact that the slender ring does not experience a torque in the $x$ or $y$ directions (i.e., a restoring torque) when the axis of symmetry aligns perfectly with the $y$-axis in the inertial frame, since the gradient of the fluid velocity is not used in the calculation of the slender body torque. Hence the ring is susceptible to wobbling/precession at this orientation. This is in contrast with the spheroid, which experiences a non-zero torque in shear flow because of the positive fluid velocity gradient, regardless of the particle orientation. These discrepancies may not appear in more complex 3D flows and geometries.

\section{Conclusion} 
In this paper we consider a model for thin, rigid fibers in viscous flows based on slender body theory. We investigate using the slender body approximation for the fluid field away from the fiber centerline as an approximation for the motion of the fiber itself by evaluating the expression on a curve along the slender body surface. Numerically, this yields a matrix equation for the force density along the length of the fiber that appears to be suitable for inversion even for very fine discretization of the fiber centerline. 

For simple geometries and simple flows, we compare the slender body model to exact or asymptotically accurate expressions for the total force and torque acting on the particle. For the thin prolate spheroid, we compare the Stokes drag force predicted by slender body theory to the exact expression of Chwang and Wu \cite{chwang1975hydromechanics}; for the cylinder, we compare with the asymptotic expression of Keller and Rubinow \cite{keller1976slender}; and for the thin torus, we compare with the asymptotic force expression of Johnson and Wu \cite{johnson1979hydromechanics}. In the case of the prolate spheroid and the thin torus, we find essentially $\mathcal{O}(\epsilon\log\epsilon)$ agreement between our model and the exact or asymptotically accurate force values (tables \ref{validate2} and \ref{tab:validate1}), which is the accuracy predicted by rigorous error analyses \cite{closed_loop,free_ends}. 

We also compared the torques on a thin prolate spheroid in shear flow for which the exact torques are given by Jeffery \cite{jeffery1922motion}. In the case of a thin torus, we qualitatively compared the dynamics of the torus with the Jeffery torques on an oblate spheroid of similar size. For the prolate spheroid, we found good agreement between our model and the Jeffery model, especially as the aspect ratio of the particle increases. In particular, in the slender body model, the dynamics appear to be better resolved for finer discretization of the filament (large $n$). For the thin torus, we observe somewhat similar results to those of Challabotla \cite{challabotla2015rotational} for oblate spheroids; namely, we observe an initial ``wobbling" phase followed by a steady ``rolling" phase. The main difference is that in the rolling phase, the thin torus precesses about the directions perpendicular to the shear plane, while the spheroid maintains a constant angular momentum. This may be due to the fact that the slender model does not explicitly experience torque through the gradient, but only the values of the fluid velocity at the location of the centerline. 

In the future, we aim to use this model to simulate elongated particles to determine the length scale at which the Jeffery model for prolate spheroids begins to lose validity in turbulent flows. We also aim to study the aggregation properties of many slender particles with more complicated shapes in turbulence (for example, helices or arbitrary closed loops). On the theoretical side, we would also like to obtain a more complete characterization of solvability conditions for the centerline equation. This would involve a spectral analysis of the equation \eqref{SB_new} as well as the slender body PDE of \cite{closed_loop,free_ends}.

\appendix
\section{Other limiting slender body velocity expressions}\label{append}
Here we provide a brief overview of other methods used to obtain an expression for the motion of the fiber centerline $\frac{\p \X(s,t)}{\p t}$. 

One such method is that of Lighthill \cite{lighthill1976flagellar} in which, away from $s=s'$, we simply plug $\bx=\X(s)$ into the integral expression \eqref{stokes_SB} (note that the doublet has negligible effect away from $s=s'$). Near $s=s'$, under the assumption that the centerline is essentially straight and the force density is approximately constant within this small region, the expression \eqref{stokes_SB} can be evaluated exactly to obtain
\begin{equation}\label{lighthill}
\begin{aligned}
8\pi\mu \big( \bu^{\rm L}(s,t)- \bu_0(\X(s,t),t)\big) &= 2({\bf I}-\be_s\be_s^{\rm T})\bm{f}(s,t) + \int_{\abs{\bm{R}_0}>\delta} \bigg(\frac{{\bf I}}{\abs{\bm{R}_0}}+ \frac{\bm{R}_0\bm{R}_0^{\rm T}}{\abs{\bm{R}_0}^3} \bigg)\bm{f}(s',t) \, ds'; \\
\bm{R}_0(s,s',t) &=\X(s,t) - \X(s',t), \quad \delta = \epsilon r(s) \sqrt{e}/2.
\end{aligned}
\end{equation}
Here $\bu^{\rm L}(s,t)$ approximates $\frac{\p \X(s,t)}{\p t}$, the actual motion of the fiber centerline, and $\bu_0(\X(s,t),t)$ is the fluid flow at the spatial point $\bx=\X(s,t)$ in the absence of the fiber. 

Another popular method is that of Keller and Rubinow \cite{keller1976slender} in which the expression \eqref{stokes_SB} is evaluated on the {\it actual} slender body surface (i.e. at a distance $\epsilon r(s)$ from $\X(s,t)$) and the method of matched asymptotics is used to obtain an expression for $\epsilon=0$. In the far field (away from $s=s'$), \eqref{stokes_SB} is simply Taylor expanded about $\epsilon=0$. In the near field (near $s=s'$), the expression \eqref{stokes_SB} is rewritten in terms of the rescaled variable $\xi = (s-s')/\epsilon$ and then expanded about $\epsilon=0$. The far- and near-field expressions are then matched to create a centerline velocity expression that includes a local operator and a singular finite-part non-local operator:
\begin{equation}\label{slender body theory_KR}
8\pi\mu \big(\bu^{\rm KR}(s,t) - \bu_0(\X(s,t),t) \big) = -\bm{\Lambda}[\bm{f}](s,t) - \bm{K}[\bm{f}](s,t). 
\end{equation}
In the free end setting, the operators $\bm{\Lambda}$ and $\bm{K}$ are given by 
 \begin{equation}\label{free_LK}
\begin{aligned}
\bm{\Lambda}[\bm{f}](s,t) &:=  \big[({\bf I}- 3\be_{\rm s}\be_{\rm s}^{\rm T}) + ({\bf I}+\be_{\rm s}\be_{\rm s}^{\rm T}) L(s) \big]{\bm f}(s,t) \\
\bm{K}[\bm{f}](s,t) &:= \int_{-1/2}^{1/2} \left[ \left(\frac{{\bf I}}{|\bm{R}_0|}+ \frac{\bm{R}_0\bm{R}_0^{\rm T}}{|\bm{R}_0|^3}\right){\bm f}(s',t) - \frac{{\bf I}+\be_{\rm s}(s)\be_{\rm s}(s)^{\rm T} }{|s-s'|} {\bm f}(s,t)\right] \, ds',
\end{aligned}
\end{equation}
where $L(s) = \log\big( \frac{2(1/4-s^2)+2\sqrt{(1/4-s^2)^2+4\epsilon^2r^2(s)}}{\epsilon^2r^2(s)}\big)$. Note that we define $L$ in this way to avoid singularities at the fiber endpoints; thus this $L$ differs slightly from the expression given by \cite{gotz2000interactions} or the expression in \cite{tornberg2004simulating}. 

In the closed loop setting, $\bm{\Lambda}$ and $\bm{K}$ are given by
\begin{equation}\label{closed_LK}
\begin{aligned}
\bm{\Lambda}[\bm{f}](s,t) &:=  \big[({\bf I}- 3\be_{\rm s}\be_{\rm s}^{\rm T})-2({\bf I}+\be_{\rm s}\be_{\rm s}^{\rm T}) \log(\pi\epsilon/4) \big]{\bm f}(s,t) \\
\bm{K}[\bm{f}](s,t) &:= \int_{\T} \left[ \left(\frac{{\bf I}}{|\bm{R}_0|}+ \frac{\bm{R}_0\bm{R}_0^{\rm T}}{|\bm{R}_0|^3}\right){\bm f}(s',t) - \frac{{\bf I}+\be_{\rm s}(s)\be_{\rm s}(s)^{\rm T} }{|\sin (\pi(s-s'))/\pi|} {\bm f}(s,t)\right] \, ds'.
\end{aligned}
\end{equation}

However, a spectral analysis of the Keller-Rubinow operator $-(\bm{\Lambda}+\bm{K})$ in the case of simple fiber geometries (see G\"otz \cite{gotz2000interactions} for the straight centerline and Shelley and Ueda \cite{shelley2000stokesian} for the circular centerline) shows that the Keller-Rubinow expression is not suitable for inversion. In particular, the operator $-(\bm{\Lambda}+\bm{K})$ has a vanishing or nearly vanishing eigenvalue at some wavenumber $k\sim1/\epsilon$. This high wavenumber instability limits the level to which the fiber can be discretized for numerics. It seems likely that more complicated centerline geometries also lead to a similar conclusion. Therefore in order to use the Keller-Rubinow expression for numerical simulations, the kernel of the operator $\bm{K}$ must be regularized. For example, in \cite{shelley2000stokesian,tornberg2004simulating}, the denominators in the kernel of $\bm{K}$ are replaced by $\sqrt{\abs{\bm{R}_0}^2+\delta^2}$ and $\sqrt{\sin^2(\pi(s-s'))/\pi^2+\delta^2}$, where $\delta=\delta(\epsilon)$ is chosen according to the fiber radius to maintain the same asymptotic accuracy as the Keller-Rubinow expression. This regularization, however, lacks a physical justification and clear connection to the expression \eqref{stokes_SB}.  

Another common technique for describing the motion of the fiber centerline is to instead use the method of regularized Stokeslets (see \cite{bouzarth2011modeling,cortez2005method,cortez2012slender}) to obtain an alternate version of \eqref{stokes_SB}. In this method, the Stokeslet is approximated by the (smooth) solution to  
\[ -\mu\Delta \bu + \nabla p = \bm{f} \phi_\delta(\bx), \quad  \div\bu=0 \]
where $\phi_\delta$ is a smooth, radially symmetric function with $\int_{\R^3}\phi_\delta=1$. The parameter $\delta$ determines the spread of $\phi_\delta$ and, in the case of slender body theory, is usually chosen such that $\delta\sim \epsilon$. The slender body approximation is then constructed as in \eqref{stokes_SB}, but now the resulting expression is not singular at $\bx=\X(s)$, and the velocity of the slender body itself may be approximated by simply evaluating the regularized expression along the fiber centerline. The method of regularized Stokeslets can be used to construct regularized versions of the Lighthill and Keller-Rubinow expressions \cite{cortez2012slender}. However, from the outset, the method of regularized Stokeslets approximates a slightly different problem from \eqref{stokes_SB}, and it is not entirely clear that these solutions should be close for any $\delta$. The choice of regularization parameter $\delta$ greatly affects the resulting dynamics; however, a systematic justification for this parameter choice is lacking.


\bibliography{NTNU_bib}

\begin{thebibliography}{10}

\bibitem{amarakoon1982drag}
A.~Amarakoon, R.~Hussey, B.~J. Good, and E.~G. Grimsal.
\newblock Drag measurements for axisymmetric motion of a torus at low
  {R}eynolds number.
\newblock {\em Phys. Fluids}, 25(9):1495--1501, 1982.

\bibitem{batchelor1970slender}
G.~Batchelor.
\newblock Slender-body theory for particles of arbitrary cross-section in
  {S}tokes flow.
\newblock {\em J. Fluid Mech.}, 44(3):419--440, 1970.

\bibitem{bouzarth2011modeling}
E.~L. Bouzarth and M.~L. Minion.
\newblock Modeling slender bodies with the method of regularized {S}tokeslets.
\newblock {\em J. Comput. Phys.}, 230(10):3929--3947, 2011.

\bibitem{challabotla2015rotational}
N.~R. Challabotla, C.~Nilsen, and H.~I. Andersson.
\newblock On rotational dynamics of inertial disks in creeping shear flow.
\newblock {\em Phys. Lett. A}, 379(3):157--162, 2015.

\bibitem{chwang1975hydromechanics}
A.~T. Chwang and T.~Y.-T. Wu.
\newblock Hydromechanics of low-{R}eynolds-number flow. {P}art 2: Singularity
  method for {S}tokes flows.
\newblock {\em J. Fluid Mech.}, 67(4):787--815, 1975.

\bibitem{cortez2005method}
R.~Cortez, L.~Fauci, and A.~Medovikov.
\newblock The method of regularized {S}tokeslets in three dimensions: analysis,
  validation, and application to helical swimming.
\newblock {\em Phys. Fluids}, 17(3):031504, 2005.

\bibitem{cortez2012slender}
R.~Cortez and M.~Nicholas.
\newblock Slender body theory for {S}tokes flows with regularized forces.
\newblock {\em Commun. Appl. Math. Comput. Sci.}, 7(1):33--62, 2012.

\bibitem{cox1970motion}
R.~Cox.
\newblock The motion of long slender bodies in a viscous fluid part 1. general
  theory.
\newblock {\em J. Fluid Mech.}, 44(4):791--810, 1970.

\bibitem{gallily1979orderly}
I.~Gallily and A.-H. Cohen.
\newblock On the orderly nature of the motion of nonspherical aerosol
  particles. ii. inertial collision between a spherical large droplet and an
  axially symmetrical elongated particle.
\newblock {\em J. Colloid Interface Sci.}, 68(2):338--356, 1979.

\bibitem{goldstein2002classical}
H.~Goldstein, C.~Poole, and J.~Safko.
\newblock Classical mechanics, 2002.

\bibitem{gotz2000interactions}
T.~G{\"o}tz.
\newblock {\em Interactions of fibers and flow: asymptotics, theory and
  numerics}.
\newblock Doctoral dissertation, University of Kaiserslautern, 2000.

\bibitem{hancock1953self}
G.~Hancock.
\newblock The self-propulsion of microscopic organisms through liquids.
\newblock {\em Proc. R. Soc. Lond. A}, 217(1128):96--121, 1953.

\bibitem{jeffery1922motion}
G.~B. Jeffery.
\newblock The motion of ellipsoidal particles immersed in a viscous fluid.
\newblock {\em Proc. R. Soc. Lond. A}, 102(715):161--179, 1922.

\bibitem{johnson1980improved}
R.~E. Johnson.
\newblock An improved slender-body theory for {S}tokes flow.
\newblock {\em J. Fluid Mech.}, 99(02):411--431, 1980.

\bibitem{johnson1979hydromechanics}
R.~E. Johnson and T.~Y. Wu.
\newblock Hydromechanics of low-{R}eynolds-number flow. {P}art 5: Motion of a
  slender torus.
\newblock {\em J. Fluid Mech.}, 95(2):263--277, 1979.

\bibitem{keller1976slender}
J.~B. Keller and S.~I. Rubinow.
\newblock Slender-body theory for slow viscous flow.
\newblock {\em J. Fluid Mech.}, 75(4):705--714, 1976.

\bibitem{lighthill1976flagellar}
J.~Lighthill.
\newblock Flagellar hydrodynamics.
\newblock {\em SIAM review}, 18(2):161--230, 1976.

\bibitem{majumdar1977axisymmetric}
S.~Majumdar and M.~O'Neill.
\newblock On axisymmetric {S}tokes flow past a torus.
\newblock {\em Z. Angew. Math. Phys.}, 28(4):541--550, 1977.

\bibitem{mao2014motion}
W.~Mao and A.~Alexeev.
\newblock Motion of spheroid particles in shear flow with inertia.
\newblock {\em J. Fluid Mech.}, 749:145--166, 2014.

\bibitem{martin2017deposition}
J.~Martin, A.~Lusher, R.~C. Thompson, and A.~Morley.
\newblock The deposition and accumulation of microplastics in marine sediments
  and bottom water from the irish continental shelf.
\newblock {\em Sci. Rep}, 7(1):10772, 2017.

\bibitem{closed_loop}
Y.~Mori, L.~Ohm, and D.~Spirn.
\newblock Theoretical justification and error analysis for slender body theory.
\newblock {\em Comm. Pure Appl. Math, to appear}, 2018.

\bibitem{free_ends}
Y.~Mori, L.~Ohm, and D.~Spirn.
\newblock Theoretical justification and error analysis for slender body theory
  with free ends.
\newblock {\em arXiv preprint arXiv:1901.11456}, 2019.

\bibitem{mortensen2008dynamics}
P.~Mortensen, H.~Andersson, J.~Gillissen, and B.~Boersma.
\newblock Dynamics of prolate ellipsoidal particles in a turbulent channel
  flow.
\newblock {\em Phys. Fluids}, 20(9):093302, 2008.

\bibitem{shelley2000stokesian}
M.~J. Shelley and T.~Ueda.
\newblock The {S}tokesian hydrodynamics of flexing, stretching filaments.
\newblock {\em Phys. D}, 146(1):221--245, 2000.

\bibitem{tapley2019novel}
B.~Tapley, E.~Celledoni, B.~Owren, and H.~I. Andersson.
\newblock A novel approach to rigid spheroid models in viscous flows using
  operator splitting methods.
\newblock {\em Numer. Algorithms}, pages 1--19, 2019.

\bibitem{tapley2019computing}
B.~K. Tapley.
\newblock Computing cost-effective particle trajectories in numerically
  calculated incompressible fluids using geometric methods.
\newblock {\em arXiv preprint arXiv:1901.05236}, 2019.

\bibitem{tornberg2004simulating}
A.-K. Tornberg and M.~J. Shelley.
\newblock Simulating the dynamics and interactions of flexible fibers in
  {S}tokes flows.
\newblock {\em J. Comput. Phys.}, 196(1):8--40, 2004.

\bibitem{voth2017anisotropic}
G.~A. Voth and A.~Soldati.
\newblock Anisotropic particles in turbulence.
\newblock {\em Annu. Rev. Fluid Mech.}, 49:249--276, 2017.

\bibitem{zhang2001ellipsoidal}
H.~Zhang, G.~Ahmadi, F.-G. Fan, and J.~B. McLaughlin.
\newblock Ellipsoidal particles transport and deposition in turbulent channel
  flows.
\newblock {\em Int. J. Multiph. Flow}, 27(6):971--1009, 2001.

\bibitem{zhao2015rotation}
L.~Zhao, N.~R. Challabotla, H.~I. Andersson, and E.~A. Variano.
\newblock Rotation of nonspherical particles in turbulent channel flow.
\newblock {\em Phys. Rev. Lett.}, 115(24):244501, 2015.

\end{thebibliography}


\end{document}